\documentclass[aip,nofootinbib]{revtex4-1}

\usepackage{graphicx}
\usepackage[colorlinks=true, allcolors=blue]{hyperref}
\usepackage{subfig}
\usepackage{amsmath}
\usepackage{siunitx}
\usepackage{caption}
\usepackage{setspace}

\begin{document}

\preprint{LA-UR-18-28957}

\title{Extreme background-rejection techniques for the ELROI optical satellite license plate}

\author{Rebecca M. Holmes}
\email[]{rmholmes@lanl.gov}
\altaffiliation{ISR-1, Space Science \& Applications}

\affiliation{Los Alamos National Laboratory, Los Alamos, NM}

\author{David M. Palmer}
\email[]{palmer@lanl.gov}
\altaffiliation{ISR-2, Space \& Remote Sensing}

\affiliation{Los Alamos National Laboratory, Los Alamos, NM}

\date{\today}

\begin{abstract}

The Extremely Low-Resource Optical Identifier (ELROI) is a concept for an autonomous, low-power optical ``license plate'' that can be attached to anything that goes into space. ELROI uses short, omnidirectional flashes of laser light to encode a unique ID number which can be read by a small ground telescope using a photon-counting sensor and innovative extreme background-rejection techniques. ELROI is smaller and lighter than a typical radio beacon, low-power enough to run on its own small solar cell, and can safely operate for the entire orbital lifetime of a satellite or debris object. The concept has been validated in ground tests, and orbital prototypes are scheduled for launch in 2018 and beyond. In this paper we focus on the details of the encoding scheme and data analysis that allow a milliwatt optical signal to be read from orbit. We describe the techniques of extreme background-rejection needed to achieve this, including spectral and temporal filtering, and discuss the requirements for an error-correcting code to encode the ID number. Worked examples with both simulated and experimental (ground test) data will illustrate the methods used. We present these techniques to describe a new photon-counting optical communication concept, and to encourage others to consider observing upcoming test flights.
\end{abstract}

\maketitle %\maketitle must follow title, authors, abstract and \pacs

%%%%%%%%%%%%%%%%%%%%%%%%%%  body  %%%%%%%%%%%%%%%%%%%%%%%%%%
\section{Introduction}
\label{introduction}
% Reference The Inter-Agency Space Debris Coordination Comittee
% 4.3.5 Trackability

%The load on surveillance systems will grow dramatically with the deployment of large constellations. Likewise, the number of conjunction events in these altitudes will grow. Enhancing %trackability, e.g. by adding onboard active and/or passive components can improve the orbit determination and prediction. This would have positive impact on conjunction analysis.

%It is recommended to enhance trackability by adding onboard active and/or passive components

\begin{figure}[h]
\centering 
\subfloat[]{\includegraphics[width=0.4\textwidth]{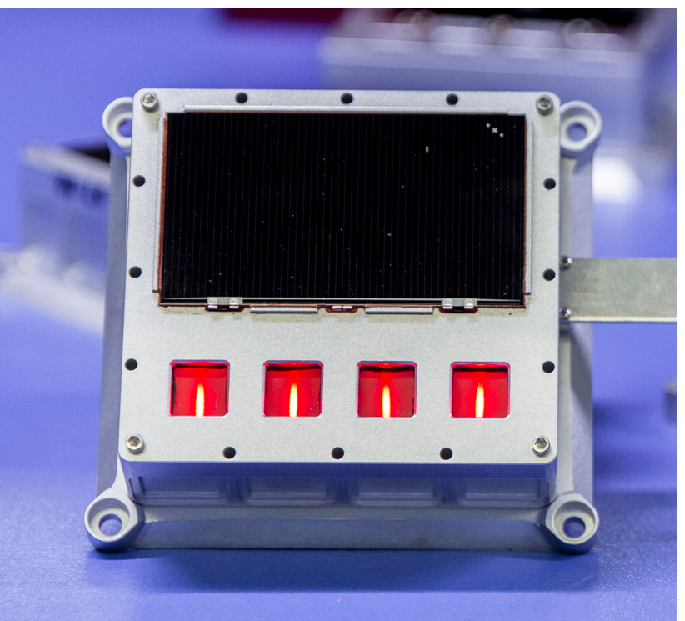}}
\hspace{0.1\textwidth}
\subfloat[]{\includegraphics[width=0.35\textwidth]{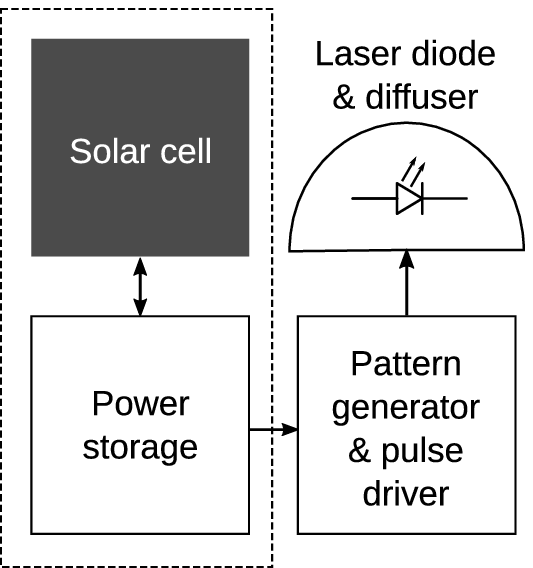}}
\caption{
    a. A current-generation ELROI prototype, showing laser diodes and solar cell. The design shown is approximately 10 cm x 10 cm x 3 cm and 300 g. A mature design, in progress, can be miniaturized to the size of a postage stamp with 0.5 cm thickness. Details may be found in \cite{Holmes2018}.
    b. Simplified block diagram of the ELROI beacon electronics.
    }
\label{fig:hardware}
\end{figure}

The Extremely Low-Resource Optical Identifier (ELROI) is a low-power optical ``license plate'' that can be attached to anything that goes into space \cite{Palmer2018,Palmer2016,Palmer2015}. ELROI is designed to help address the problem of space object identification (SOI) in the crowded space around the Earth, where over 19,000 objects---from active satellites to debris fragments---are currently tracked and monitored \cite{spacetrack}. Tracking these objects is a multi-billion USD effort, and requires continuous knowledge of each object's position and trajectory. Sudden orbit changes or interruption in tracking can lead to confusion about a tracked object's identity, and multiple costly observations may be needed to re-identify it as a known satellite. SOI is easier if a satellite carries a continuous identifying beacon that can be read by anyone on the ground, but no suitable standard technology is currently used for this purpose \cite{IADC2017,Peterson2018}. ELROI is a new optical SOI technology which is currently at the prototype stage (Figure \ref{fig:hardware}). The concept has been validated in ground tests \cite{Palmer2018}, and orbital prototypes are scheduled for launch in late 2018 and beyond. 

ELROI is an autonomous solar-powered optical beacon that uses short, diffused flashes of laser light at milliwatt average power to encode a unique ID number, which can be read by a small ground telescope during local night using a photon-counting sensor and innovative extreme background-rejection techniques. ELROI is smaller and lighter than a typical radio beacon, suitable for small satellites including CubeSats, and can safely operate for the entire orbital lifetime of a satellite without concern for radio-frequency interference (RFI). Autonomous solar power also allows ELROI to continue emitting after the end of a satellite's operational lifetime, and makes ELROI suitable for non-powered debris objects such as rocket bodies. Space object identification is generally more challenging than tracking with current technology, and ELROI is designed to address the problem of objects that are tracked but not identified; thus, reading the ELROI ID number requires tracking information for the host satellite (such as the publicly available and regularly updated data in the SpaceTrack catalog \cite{spacetrack}).

This paper is not a comprehensive overview of the ELROI system, and additional details about the beacon hardware, ground station designs, link budgets, and the concept of operations may be found in \cite{Palmer2018}. Here, we focus on the details of the encoding scheme and data analysis that allow a milliwatt optical signal to be read from orbit. We will describe the techniques of extreme background-rejection needed to achieve this, including spectral filtering and temporal filtering using a period- and phase-recovery algorithm, and discuss the requirements for an error-correcting code to encode the ID number. Worked examples with both simulated and experimental (long-range ground test) data will illustrate the methods used.

We encourage others to consider observing our test flights. Details about current flight prototypes, as well as practical considerations and necessary equipment for those interested in observing ELROI test flights from their own ground stations, may be found in \cite{Holmes2018}.

\subsection{Characteristics of the ELROI signal}
\label{signal-characteristics}

\begin{figure}[h]
\centering
\includegraphics[width=0.9\textwidth]{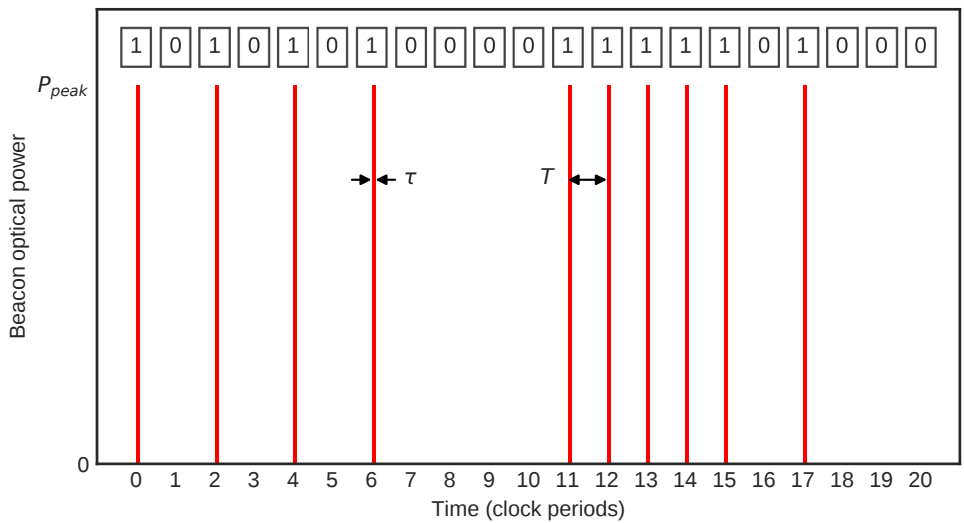}
\caption{An example of the signal produced by an ELROI beacon. The onboard laser diode emits short pulses of light (pulse width~=~$\tau$) separated by a fixed period (clock period~=~$T$). Each clock period encodes one bit of the beacon ID number (shown at top): if the bit is a 1, the beacon emits a pulse, and if the bit is a 0, the beacon does not emit a pulse. The laser power during a pulse is $P_{\textrm{peak}}$; otherwise it is zero. The pulse width in this example is exaggerated compared to the clock period. For a real beacon, a typical value of $\tau / T$ is $1/1000$. The ID number repeats many times per second.}
\label{fig:simple-signal}
\end{figure}

ELROI uses short, bright flashes of laser light at a fixed period to encode a unique binary ID number with an on-off keying (OOK) scheme (Figure \ref{fig:simple-signal}). The approach builds on photon-counting laser optical communications techniques \cite{Farrell2018a,Farrell2018b,Mendenhall2007,Boroson2004}, but in a novel operating regime at much lower power, very low bandwidth, and without strict pointing requirements due to the very broad optical beam. The ID number repeats many times per second. The laser light is nearly monochromatic, with a spectral range of about 1~nm. The peak power of the laser is high ($P_{\textrm{peak}} \approx$ 1~watt), but it is pulsed at a low duty cycle so that the average power is reduced to a few milliwatts.

This low average power requirement is essential for autonomous solar-powered operation in space, and allows the beacon to be ``low-resource'' in size and mass. Light from the beacon is diffused over a wide angle---approximately 145 degrees for current prototypes---to increase visibility from the ground without active pointing. At orbital distances, this combination of low optical power and large solid angle results in an extremely weak signal at a ground station. Link budget calculations for a typical design indicate that a ground station will detect 3-4 signal photons/s from a 2-mW ELROI beacon in low-Earth orbit (LEO) at 1000 km range \cite{Palmer2018}. This is far below the 90 photons/s of background light (at the beacon wavelength) expected from a small sunlit CubeSat host. However, the spectral and timing characteristics of the signal enable extreme background-rejection techniques, introduced in Section \ref{ebr}, which allow the beacon ID to be read reliably after a few minutes of accumulating data---i.e., in a single pass over a ground station for a typical LEO orbit. 

Reading an ELROI ID requires a photon-counting sensor, such as a single-photon avalanche diode (SPAD) or photomultiplier. The sensor must, at minimum, register discrete signals for each detected photon with timing resolution better than the pulse width $\tau$. (Due to read noise, conventional CCD or CMOS detectors are not suitable.) The sensor must receive light from a telescope, which may be relatively small, but must be able to accurately and precisely track objects across the sky. An imaging or position-sensitive detector is useful to reduce the pointing and tracking tolerance requirements on the ground station telescope, and our current ground station uses a LANL-developed photon-counting camera which combines a photocathode, microchannel plate (MCP), and multi-anode readout \cite{Thompson2013}. However, an imaging sensor is not required to read the ELROI ID. Further discussion of the design trade-offs between sensor, ground station, and beacon may be found in \cite{Palmer2018}~and~\cite{Holmes2018}.

The data recorded from the ground station is a list of photon detection times. Some of these times correspond to signal photons, and many more are background photons, either from the environment or from noise in the sensor. The goal of data analysis is to isolate enough of the signal photons to determine the value of each ID bit with sufficient confidence to recover the ID number. Each step in this analysis can be done efficiently, so that a satellite can be identified in real time as it is tracked over the ground station.

\subsection{Background rejection}
\label{ebr}
% This section should reference other similar approaches to optical comm, JPL and SETI

The critical characteristics of the ELROI signal that enable extreme background rejection are the narrow spectral range of the laser light, the precise timing of the signal clock period, and the high peak power and narrow width of the signal pulses relative to the clock period.

\subsubsection{Spectral filtering}
\label{spectral}
The narrow spectral range of the beacon light allows most environmental background photons to be blocked with a narrow-band optical filter. Current prototypes use 638-nm laser diodes with a spectral width of less than 1~nm. (This wavelength was chosen to match the sensitivity of the LANL photon-counting camera; other wavelengths, particularly near-IR, may be more appropriate for future designs.) For an orbital beacon, the majority of background photons are sunlight reflected from the host satellite, with some sky background depending on environmental conditions (particularly moon phase) and tracking accuracy. A 10-nm filter centered at 638 nm blocks about 99\% of reflected sunlight. The filter bandwidth is a compromise between background transmission and tolerance for thermal shifts in the central wavelength of the beacon. Spectral filtering is an important part of the ELROI system, but the rest of this paper will focus on background rejection in data analysis after spectrally filtered photons have been detected. All background photon rates will be assumed to be measured after appropriate spectral filtering.

\subsubsection{Phase cut}
\label{phase-cut}

\begin{figure}[ht!]
\centering
\includegraphics[width=0.8\textwidth]{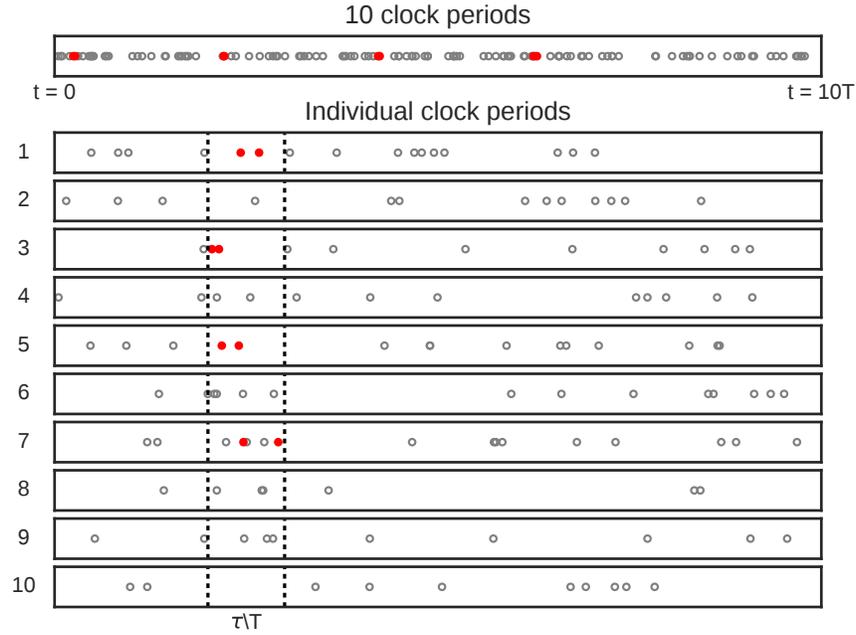}
\caption{A simple example of the phase cut used to reject background photons. The top panel shows simulated photon detection times from a sensor observing the beacon that was introduced in Figure \ref{fig:simple-signal}, with signal photons shown in red (filled) and background photons in gray (unfilled). Time increases from left to right and the observation covers 10 clock periods. The average background detection rate is about 10 times the average signal detection rate. Below, the same 10 clock periods are expanded and stacked to demonstrate that all signal photons arrive within a phase window with width $\tau/T$. Photons outside this window can be rejected as background. As in Figure \ref{fig:simple-signal}, the ratio $\tau/T$ is shown much larger here than for a real signal. The signal photon rate is also exaggerated compared to a real beacon, and in a real application, the beacon must be observed for several minutes spanning many repetitions of the ID number.}
\label{fig:phase-cut}
\end{figure}

Because the ELROI signal consists of short, bright pulses that are clocked at a fixed period, all the signal photons are detected within short, precisely spaced time windows (recall Figure \ref{fig:simple-signal}). The phase of signal photons in the range of [0,1) clock cycles is restricted to a narrow peak with width $\tau / T$ (Figure \ref{fig:phase-cut}). In contrast, background photons are detected at random times. Thus, if the clock period is known, a phase cut can be used to reject any photons outside this narrow phase peak. This is analogous to ``lock-in'' detection methods. The phase cut reduces background counts by a factor of $\tau / T$, which is between 1/1000 and 2/500 for typical signals. This is the advantage of using short pulses relative to the clock period.  % Note that when analyzing real data, the phase of the signal photons relative to the start of the clock period windows is unknown, and sufficient signal photons must be accumulated to show a clear phase peak before the phase cut can be applied.

Ignoring out-of-phase data is common in optical communications, but the ELROI data analysis requires much more accurate knowledge of the clock period than typical high-power applications where transitions between 0 and 1 bits can be used to continuously synchronize the receiver clock to the emitter. In the simple example of Section \ref{example-sim}, we will assume that the clock period is known to arbitrary precision. In practice, even if the nominal clock period is known, temperature variations and the tolerance of the beacon on-board clock will lead to small deviations from the nominal value, and even drift over time. To achieve full sensitivity, the true clock period must be recovered from the data to an accuracy of a few parts per million. This is possible for ELROI because an entire 2-3 minute observation can be used to determine the clock period in post-processing, with latency still short compared to the observation length. Efficient methods for clock detection and recovery are discussed in the more realistic example of Section \ref{example-hrt}.

\subsection{Decoding the ID number}
\label{decoding}

A typical ELROI beacon will repeat its ID number 5-10 times per second, so an observation spanning several minutes contains many repetitions of the ID number. The data from each repetition must be combined to get the best estimate of the ID bit values. Thus, after out-of-phase background photons are removed with a phase cut, the remaining photon detection times are ``folded'' by the known ID length: the time data is divided into windows the length of the ID (the number of bits $m \times T$), these windows are summed, and the result is binned by the clock period to get the total number of photons corresponding to each expected ID bit.

These photon detection totals must then be used to assign a 1 or 0 value to each bit to recover the ID. The bit value assignment must account for the fact that some background photons will remain even after the phase cut, so 0 bits may contain photons. The number of photons detected for 1 bits will also fluctuate significantly with Poisson statistics (shot noise), due to the very low signal count rate from orbit and the limited time available to accumulate photons during one pass over a ground station. Background is expected to be independent and identically distributed (i.i.d.) over short times. In most cases, the photon numbers of 1 and 0 bits will form clearly separate distributions, and a simple threshold procedure can be used to separate them (hard decision decoding). The hard decision method can also be extended to cases with poorer signal-to-noise ratios. The examples in this paper will use a hard decision method, but other possibilities are briefly discussed in Section \ref{conclusion}.

\subsubsection*{Error-checking and correction codes}
\label{ecc-intro}

Particularly for observations with very few signal photons, it is expected that some bits will be assigned the wrong value, and the recovered ID number may not exactly match the true ID. To provide tolerance for incorrect bits, an error-correcting code is used to generate the ID numbers. Error-checking and correction (ECC) codes are a proven strategy in modern digital communications, and can provide codeword error rates (the probability of incorrectly identifying an ID) orders of magnitude lower than the bit error rate (the probability of incorrectly identifying a bit). The error-correcting code used for ELROI restricts the ID numbers to a minimum Hamming distance from any other ID, so that up to some threshold number of bit errors, a single true ID will be significantly closer to the recovered ID than any other possible ID\cite{Hamming1950}. Thus, the final step in reading an ELROI ID is to search the database of all active ID numbers for the ID with the fewest number of errors relative to the recovered ID. 

For the examples in this paper, we use constant-weight 64-of-128-bit codes allowing cyclic permutation: the ID numbers are 128 bits long, 64 bits are 1 and 64 bits are 0, and two IDs that differ only by moving some bits from the start of the ID to the end of the ID are considered to be the same ID. The possible IDs are chosen such that any pair of IDs differs by at least 24 bits, which allows for millions of unique IDs while guaranteeing that up to 12 incorrect bits can be corrected. Because the start bit of the recovered ID is not known, it must be compared to each ID in the database with each of $m$ shifts, where $m$ is the number of bits in the code; however, the search can still be done quickly compared to the length of the observation. (While this example ECC scheme does not indicate the start bit of the ID number, other schemes can do so.) This scheme is sufficient to demonstrate the value of error-correcting codes for ELROI, and will be used on the current generation of orbital prototypes. There are several options for refining the final ECC code strategy for ELROI, but these are beyond the scope of this paper and are summarized only briefly in Section \ref{conclusion}.

\section{Example: Decoding simulated data}
\label{example-sim}

\begin{table}
\centering
\caption{Parameters for the simulated dataset analyzed in Section \ref{example-sim}. The 128-bit binary ID number is represented in hexadecimal for compactness.}
\label{table:sim-data}
\begin{tabular}{lll}
\hline
Parameter                       &                       & Value                              \\ \hline
ID number (hexadecimal)         &                       & 0x8345f3ca6ca6f0e338f5d598e525a912 \\
ID length                       & $m$                   & 128 bits                           \\
Pulse width                     & $\tau$                & 1 $\mu$s                           \\
Clock period                    & $T$                   & 1 ms                               \\
Average signal photon rate      & $R_s$                 & 5 photons/s (detected)                   \\
Average background photon rate  & $R_b$                 & 100 photon/s (detected)                 \\
Clock phase                     & $\phi_0$              & 0.50 cycles                        \\
ID start bit                    & $s$                   & 10                                 \\
Dataset length                  & $t_{\textrm{obs}}$    & 180 s                              \\ \hline
\end{tabular}
\end{table}

This example will analyze a simulated ELROI dataset with signal and background photon rates comparable to a beacon in LEO. The simulation uses typical beacon signal characteristics (Table \ref{table:sim-data}), and models a detector similar to the LANL photon-counting camera.
% Python code for generating simulated ELROI datasets, as well as tools and examples for analyzing both simulated and experimental data, are available at \url{http://github.com/XX}.

A three-minute simulated dataset was generated with a Monte Carlo method. The data is a list of simulated photon detection timestamps, and contains $N_{\textrm{tot}}=19,038$ timestamps total, corresponding to $N_{\textrm{s}} = 931$ signal photons from a beacon (spread over 180,000 pulses) and $N_{\textrm{bg}} = 18,107$ random background photons. The background photon rate is assumed to be detected after spectral filtering. Photon detection timestamps were generated to a resolution of 1 ns.

\subsection{Phase cut}
\label{sim-phase-cut}

\begin{figure}[h]
\centering
\includegraphics[width=0.75\textwidth]{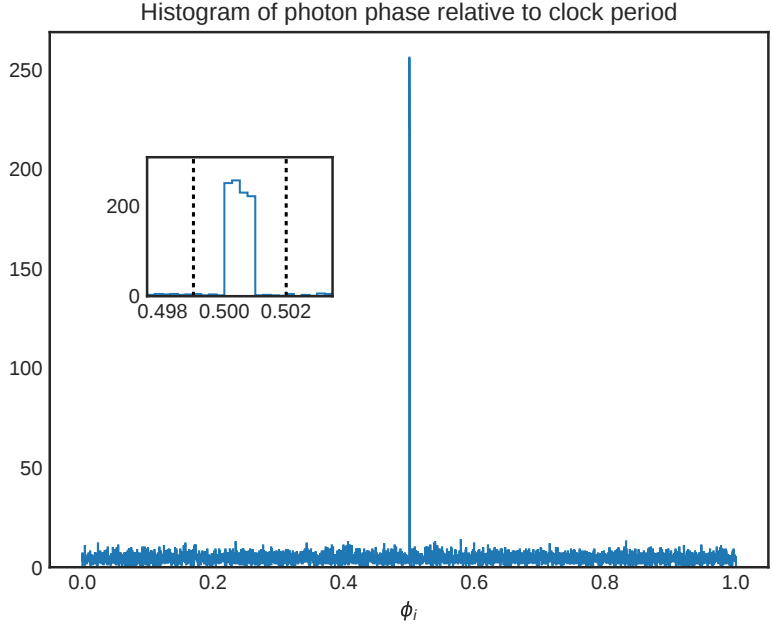}
\caption{(Simulated data) Histogram of the fractional phase $\phi_i$ of photon detection times in the range [0,1) clock cycles. Because the clock period is known exactly, the histogram shows a clear peak with width $\tau/T = 0.001$. The inset shows a closer view of the peak (each phase bin is 0.25 $\times$ $\tau/T$). The dashed lines in the inset show the phase cut window that will be applied to this data, with a width 3 $\tau/T$.}
\label{fig:sim-phase-cut}
\end{figure}

To apply the phase cut, we first compute the fractional phase $\phi_i$ of each photon in the dataset, relative to the clock period:
\begin{equation}
\label{eq:frac-phase}
\phi_i = \textrm{frac}(t_i/T)
\end{equation}
where $t_i$ is the time at which the photon was detected, $T$ is the clock period, and 
\begin{equation}
\textrm{frac}(x)=x-\lfloor {x} \rfloor
\end{equation}
is the fractional part function. $\phi_i$ is in the range [0,1) and has units of cycles. (In this example we have assumed the clock period $T$ is known exactly. The next example in Section \ref{example-hrt} will demonstrate the more realistic process of determining the clock period from the data.)

A histogram of $\phi_i$ is shown in Figure \ref{fig:sim-phase-cut}. The peak at $\phi_i = 0.5$ cycles is due to the in-phase signal photons, while the background photons have random phases. (Note that in this example, the phase of the signal photons was fixed as a simulation parameter, but in general it could take any value between 0 and 1.) We will apply a phase cut of width 3 $\tau/T$ centered on this peak, and discard all photons outside the range of 0.499 cycles to 0.502 cycles. It is clear from the histogram that a narrower phase cut would be possible without losing any signal photons, but it is not necessary for this example. For automated processing of real data, the peak identification and phase cut is done by algorithm; this will be discussed in Section \ref{example-hrt}. 

After applying the phase cut, we are left with 981 photons, of which 931 are the original signal photons and 50 are remaining background photons. Thus, the number of background photons in the dataset has been reduced to 50/18,107 = 0.28\% of the original number, without losing any of the signal photons. With this much more favorable signal-to-background ratio, we can now decode the ID number.

\subsection{Decoding the ID number}
\label{sim-decoding}

%\begin{figure}[h]
%\centering
%\includegraphics[width=0.6\textwidth]{img/sim-first-20.eps}
%\caption{(Simulated data) Photon counts for the first 20 (of $m = 128$) bits of the phase-cut data, and the binary values (top) that will be assigned to each bit in the hard decision decoding step. Note that these 20 bits of the recovered ID do not necessarily correspond to the first 20 bits of the registry ID, and the correct shift between the recovered ID and the registry ID will need to be determined.}
%\label{fig:sim-first-20}
%\end{figure}

To decode the ID number, the photon detections from each recorded repetition of the ID number must be combined. This is done by folding the remaining photon detection time data by the ID length:
\begin{equation}
t'_i = t_i \pmod{mT}
\end{equation}
where $m = 128$ is the number of bits in the ID.  Then the folded times $t'_i$ are binned into 128 clock period $T$ windows to determine the number of photons $N_{j'}$ detected for each bit $b_{j'}$ of the recovered ID number:
\begin{equation}
N_{j'} = \sum_{i} \delta_{ij} \textrm{      where     } 
\delta_{ij} =  \begin{cases}
	 1 & \text{if } j' T < t'_i \leq (j' + 1)T \\
	 0 & \text{otherwise}
 \end{cases}
\end{equation}
where $0 \leq j' < m$ and the prime indicates that we do not yet know the shift $s$ between the first bit $b_0'$ of the recovered ID number and the first bit $b_0$ of the registry ID number. %Figure~\ref{fig:sim-first-20} shows the values of $N_{j'}$ for the first 20 bits recovered from our simulated data.

\begin{figure}[h]
\centering 
\subfloat[]{\includegraphics[width=0.50\textwidth]{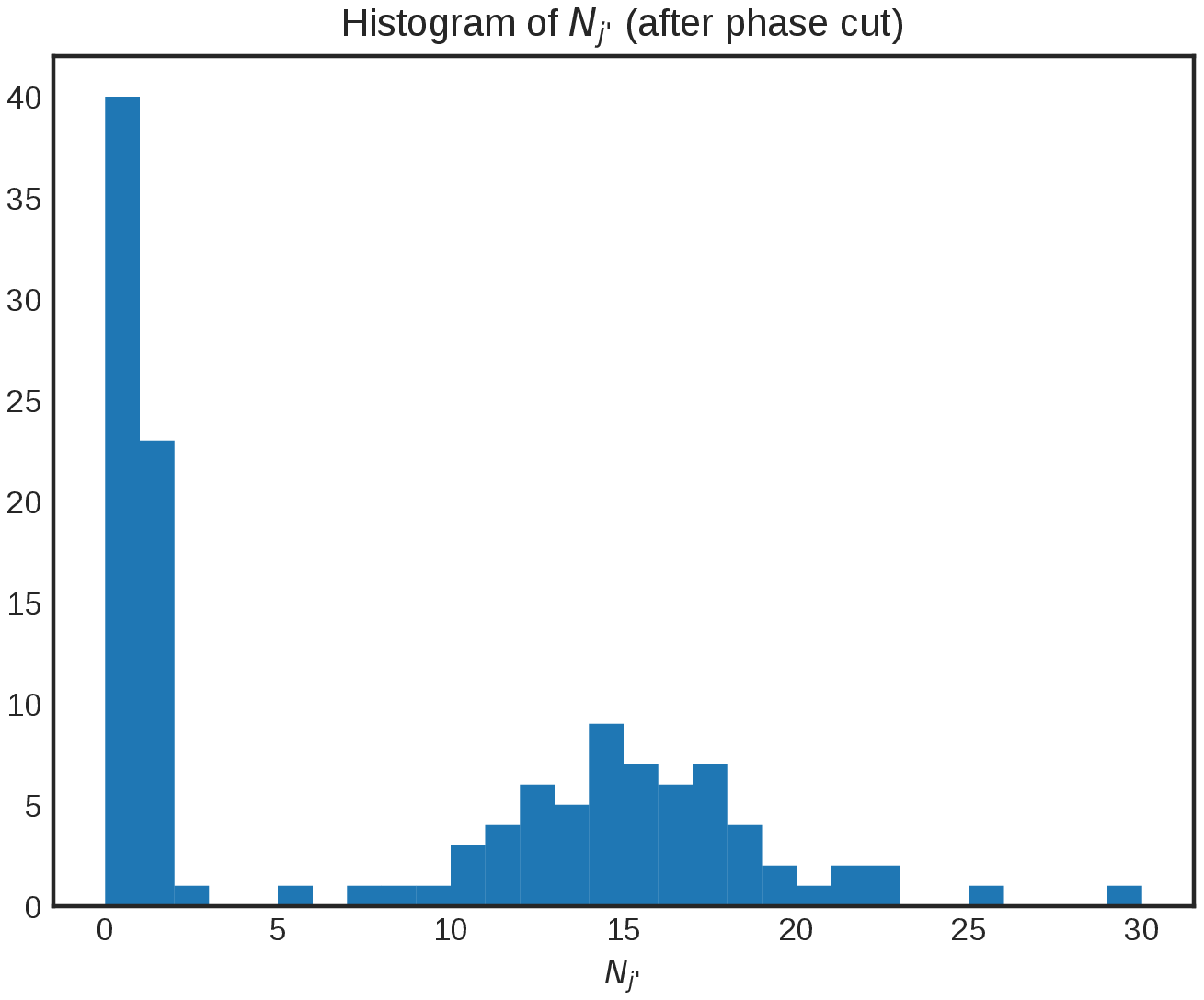}}
\subfloat[]{\includegraphics[width=0.50\textwidth]{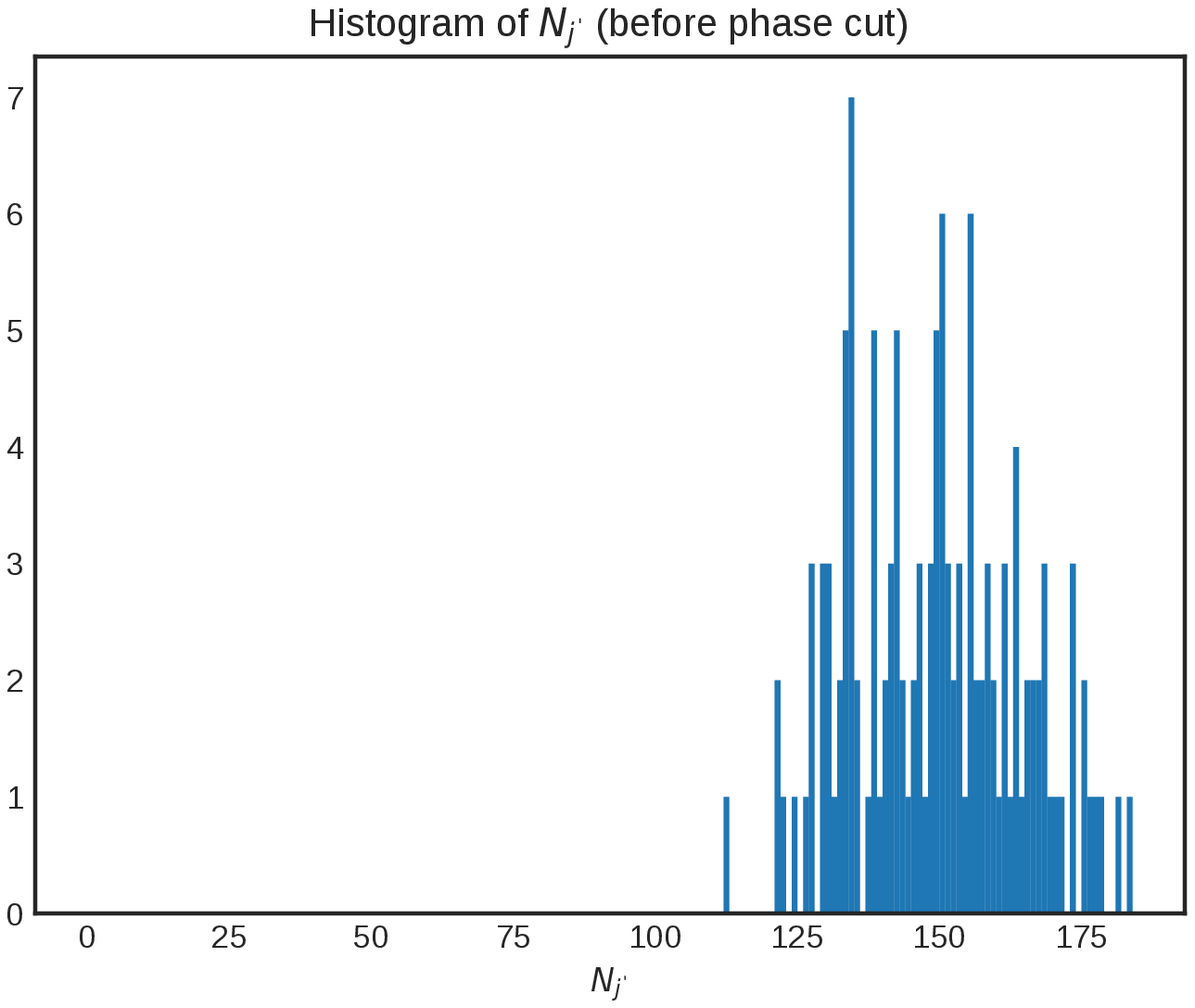}}
\caption{(Simulated data)
    a. Histogram showing the frequency of photon numbers per bit (for 981 photons in 128 bits) after the phase cut. The histogram clearly shows two distributions with different means, corresponding to 0 bits and 1 bits.
    b. Histogram showing the frequency of photon numbers per bit (for 19,038 photons in 128 bits) before the phase cut is applied to remove background photons. Note that without the phase cut, the distributions of 0 bits and 1 bits are indistinguishable.}
\label{fig:sim-bit-counts}
\end{figure}

We now assign a binary value to each bit, using a simple hard decision decoding method determined by the number of photons $N_{j'}$ in a bit:
\begin{equation}
 b_{j^\prime} = 
 \begin{cases}
	 1 & \text{if } N_{j^\prime} \geq N_{\textrm{thresh}}\\
	 0 & \text{otherwise}
 \end{cases}
\end{equation}
A histogram\footnote{Histograms are plotted with the left bin edge labeled, and bins are left-closed, right-open intervals.} of $N_{j^\prime}$ is useful for visualizing where this threshold should be set (Figure \ref{fig:sim-bit-counts}). In this example, the phase-cut data shows two clearly different distributions of photon numbers, corresponding to 0 bits and 1 bits. A threshold of $N_{\textrm{thresh}} = 5$ appears reasonable based on the histogram of $N_{j^\prime}$, and also produces equal numbers of 0 bits and 1 bits, which matches our expectation for a 64-of-128-bit code (as discussed in the next example of Section \ref{example-hrt}, this property of the ID numbers can be used to choose an appropriate value of $N_{\textrm{thresh}}$ when the distinction between 1 and 0 bits is less obvious).

%\begin{table}
%\centering
%\caption{(Simulated data) Bit errors in the recovered ID number compared to 20 possible registry ID numbers generated with a 64-of-128-bit error-correcting code. The bit errors shown are for the shift $s$ that produced the minimum number of errors. ID number 16 is the ID used to generate the simulated data.}
%\label{table:sim-bit-errors}
%\begin{tabular}{l|llllllllllllllllllll}
%Registry ID & 1  & 2  & 3  & 4  & 5  & 6  & 7  & 8  & 9  & 10 & 11 & 12 & 13 & 14 & 15 & \bf{16} & 17 & 18 & 19 & 20 \\ \hline
%Bit errors  & 53 & 52 & 52 & 46 & 48 & 48 & 50 & 50 & 52 & 54 & 52 & 50 & 50 & 48 & 44 & \bf{0}  & 52 & 48 & 52 & 52
%\end{tabular}
%\end{table}

\begin{figure}[h]
\centering
\includegraphics[width=0.75\textwidth]{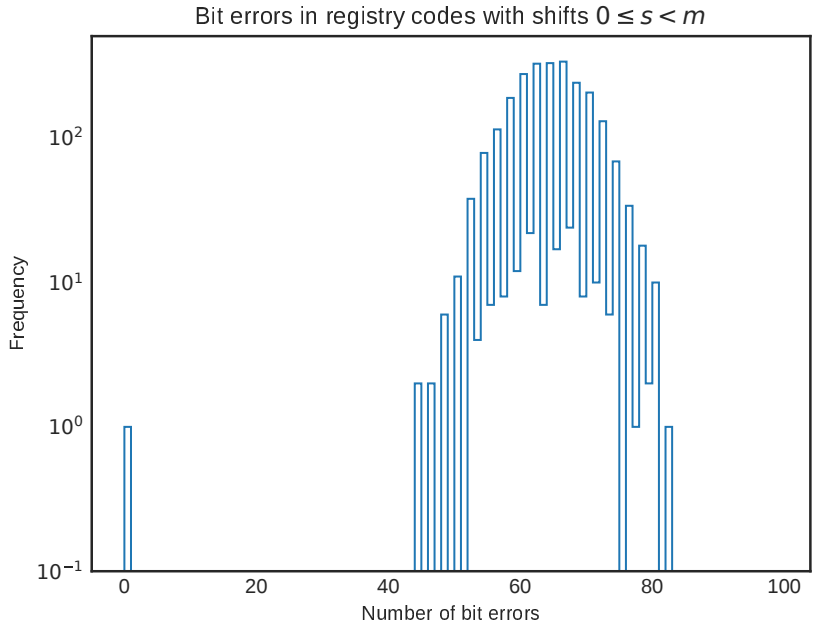}
\caption{(Simulated data) Histogram of the number of bit errors in each of 20 registry IDs with each shift $0 \leq s < m$ (2560 shifted IDs total), compared to the recovered ID. The single ID with zero errors is the correct ID.}
\label{fig:sim-bit-errors}
\end{figure}

%\begin{figure}[h]
%\centering
%\includegraphics[width=1\textwidth]{img/sim-ids.pdf}
%\caption{(Simulated data) Visualization of the recovered ID number compared to registry ID number 16. Dark bands represent 1 bits and white bands represent 0 bits. The recovered ID begins at bit 10 of the registry ID, indicated by the dashed line.}
%\label{fig:sim-ids}
%\end{figure}

The final step is to compare the recovered ID number to the registry of known IDs (with each shift $0 \leq s < m$) to find the best match. In this example we will simply find the matching ID with the smallest Hamming distance (number of bit errors) from the recovered ID. Because the ID numbers are generated with an error-correcting code, if all other IDs have a larger Hamming distance than the best match, this indicates with high confidence that our identification is correct. For this example, we will compare the recovered code to a test registry containing 20 ID numbers generated under the restrictions described in Section \ref{decoding}. (An operational registry would contain at least as many ID numbers as there are satellites with ELROI beacons, and millions of unique IDs are possible.) The result of this comparison is shown in Figure \ref{fig:sim-bit-errors}. In this simulation, we were able to get a perfect match (zero bit errors) with a registry ID %(Figure \ref{fig:sim-ids})
. The recovered ID matches registry ID number 16 with a shift $s = 10$.

\section{Example: Decoding experimental data}
\label{example-hrt}

\begin{figure}[h]
\centering
\includegraphics[width=0.75\textwidth]{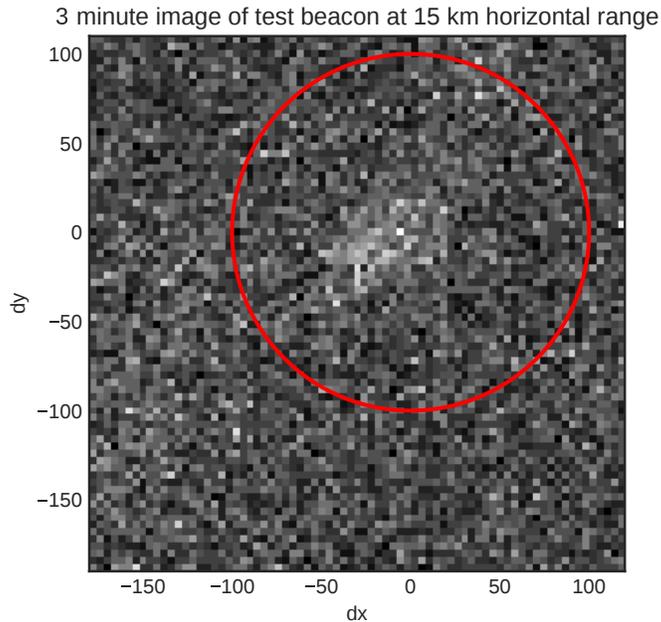}
\caption{Single-photon image of an ELROI test beacon at a horizontal range of approximately 15 km. The image represents a 3-minute accumulation with the LANL photon-counting camera, and has been spatially binned, as well as cropped to show only the area of interest. The circular aperture shows the region of the camera data (within a radius of 100 image units around the apparent location of the beacon) that was used in the analysis  of Section \ref{example-hrt}.}
\label{fig:hrt-im}
\end{figure}

% Copied (and slightly edited) from AIAA paper, may need to be revised more
The ELROI concept has been demonstrated in a ground test over approximately 15 km. This test served both to validate link budget calculations, as discussed in more detail in \cite{Palmer2018}, and to verify data analysis techniques for conditions similar to a LEO system. According to our link budget calculations, signal rates of 3 to 20 photons/s will be typical at our ground station for prototype ELROI beacons on LEO satellites, depending on the beacon power and range\cite{Palmer2018,Holmes2018}. To approximate the expected LEO link budgets, the ground test used laser diodes with lower peak power than the orbital beacons, a neutral density filter on one of the beacons to further reduce the outgoing light, and a smaller lens at a variety of apertures to replace the telescope that would be used for an orbital beacon. This allowed us to test reception at signal rates that were both higher and lower than expected for the nominal orbital case.

In this example, which uses a three-minute portion of ground test data with approximately 4.8 photons/second signal rate, we will illustrate more realistic elements of data processing, including determining the true clock period from the data, and using the known properties of the ID number to choose an appropriate bit value threshold for decoding.

\subsubsection*{ELROI ground test conditions}
The ground test used a receiver stationed 15 km away from two beacon units. Each beacon consisted of a red laser diode (638~nm), optical diffuser, and driver electronics. The lasers were driven in a 64-of-128 pattern ID with a pulse duration $\tau=\SI{2}{\micro\second}$ and a nominal clock period $T_{\textrm{nom}}=\SI{500}{\micro\second}$ (signal characteristics are summarized in Table \ref{table:hrt-data}). The two beacon units encoded different IDs. The time-averaged effective isotropic radiated power (EIRP) in the direction of the receiver was measured to be 0.3 mW, corresponding to a peak EIRP of 150 mW after adjusting for the 1:500 duty fraction. Unit 2 was attenuated by a 13\%-transmission neutral density filter to 0.04 mW EIRP, while Unit 1 was left unfiltered. This example analyzes data from Unit 2.

The receiver consisted of a LANL-developed photon-counting camera \cite{Priedhorsky2005,Thompson2013} with an f = 400 mm lens and adjustable aperture (up to f/2.8 = 143 mm diameter). The camera has a quantum efficiency of about 3.9\% at 638~nm, sub-nanosecond timing resolution, a dark count rate of a few thousand counts/second over the entire sensor, and it can detect approximately 500,000 photons/s. The receiver was equipped with a 10-nm bandpass filter centered on the transmitter laser wavelength to reduce background. The observations were made at night. Along with the adjustable sensor aperture, the beacon power and the range to the receiver were chosen to simulate predicted photon rates from typical ELROI beacons with a range of optical powers and orbital distances. Detailed link budget calculations may be found in Ref. \citenum{Palmer2018}. Radius cuts were applied to the camera data to isolate the photon detections from each beacon (Figure \ref{fig:hrt-im}) and produce a list of photon detection times to analyze. Like the simulated data in Section \ref{example-sim}, the radius cut also contains background photons, both from the environment and from the dark count rate of the sensor.

\begin{table}
\centering
\caption{Parameters for the ground test dataset analyzed in Section \ref{example-hrt}. We consider only one of the two test beacons, Unit 2. The 128-bit binary ID number is represented in hexadecimal for compactness.}
\label{table:hrt-data}
\begin{tabular}{lll}
\hline
Parameter                       &                       & Value                              \\ \hline
ID number (hexadecimal)         &                       & 0x65b0278a7cad7b5c766f056a470f01cc \\
ID length                       & $m$                   & 128 bits                           \\
Pulse width                     & $\tau$                & 2 $\mu$s                           \\
Nominal clock period            & $T_{\textrm{nom}}$    & 500 $\mu$s                         \\
Sensor aperture                 &                       & f/2.8 = 143 mm                     \\
Measured signal photon rate     & $R_s$                 & 4.8 photons/s (detected)           \\
Measured background photon rate & $R_b$                 & 50.7 photons/s (detected)          \\ 
Dataset length                  & $t_{\textrm{obs}}$    & 180 s                              \\ \hline
\end{tabular}
\end{table}

\subsection{Recovering the true clock period}

The nominal clock period of the ground test beacon is known (and programmed into the pulse generator that drives the laser), but in contrast with the the simulated data in Section \ref{example-sim}, we expect to observe some deviation from the nominal value. The fractional error tolerance $e_T$ of the beacon onboard clock determines the expected uncertainty range of $T$: $T_{\textrm{nom}} (1-e_T) < T < T_{\textrm{nom}} (1+e_T)$. For inexpensive crystal oscillators commonly used in integrated circuits, $e_T$ = 50--100 ppm is a typical value. Temperature changes, electromagnetic fields, radiation damage, and mechanical stress can also cause small deviations from the nominal period. As illustrated in Figure \ref{fig:hrt-phase-cut}, even a small deviation from the nominal period destroys the stable fractional phase relationship between signal photons that is critical for the background-rejection phase cut. Therefore, it is necessary to determine the true clock period from the photon detection data before calculating $\phi_i$.

\subsubsection{The Fast Folding Algorithm}
Given the nominal clock period and the expected value of $e_T$, the true clock period can be recovered with computationally efficient techniques such as the Fast Folding Algorithm (FFA) \cite{Staelin1969}. Details on the FFA may be found in \citenum{Staelin1969}; what follows is a brief conceptual summary.

\begin{figure}[h]
\centering 
\subfloat[]{\includegraphics[width=0.5\textwidth]{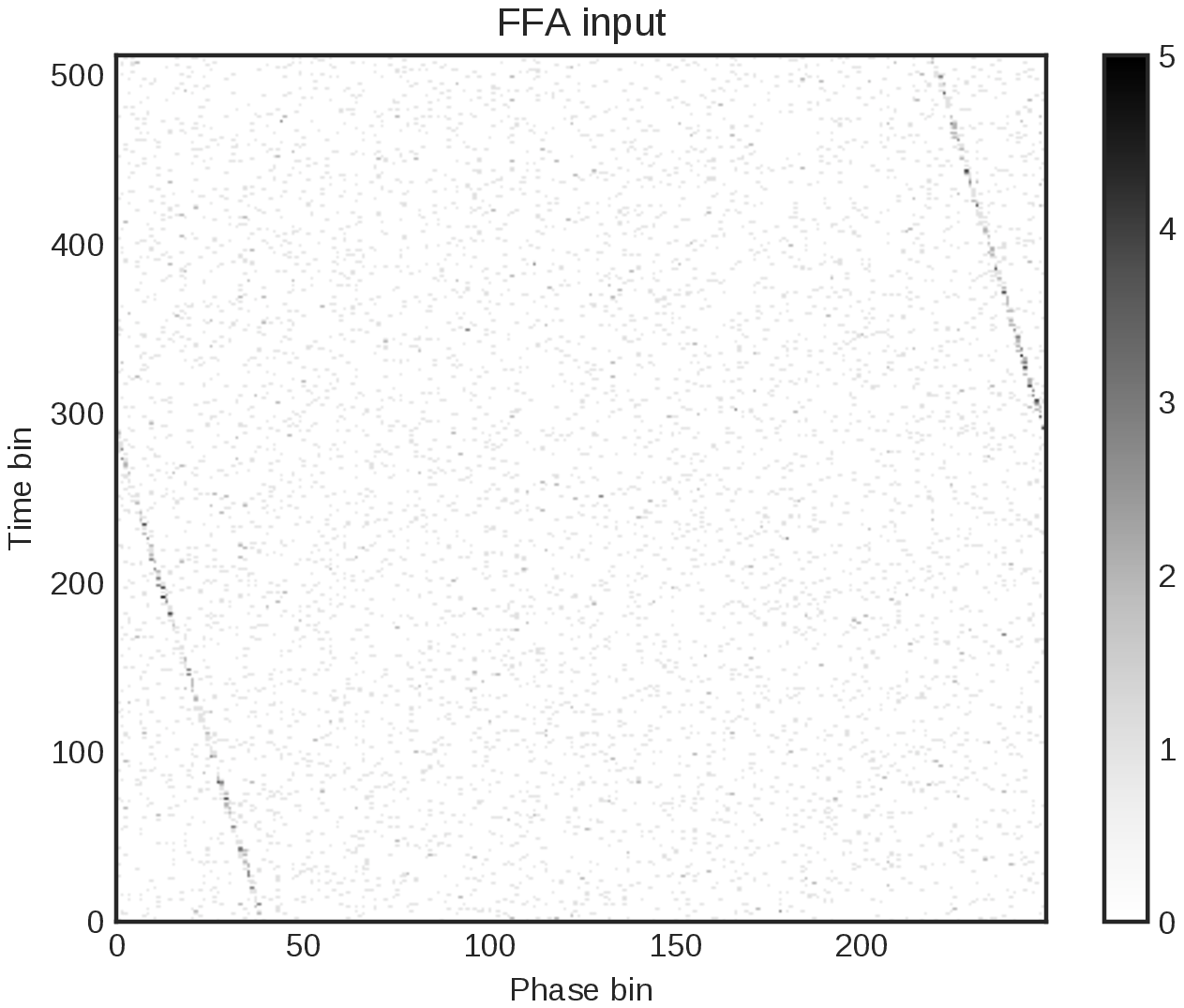}}
\subfloat[]{\includegraphics[width=0.5\textwidth]{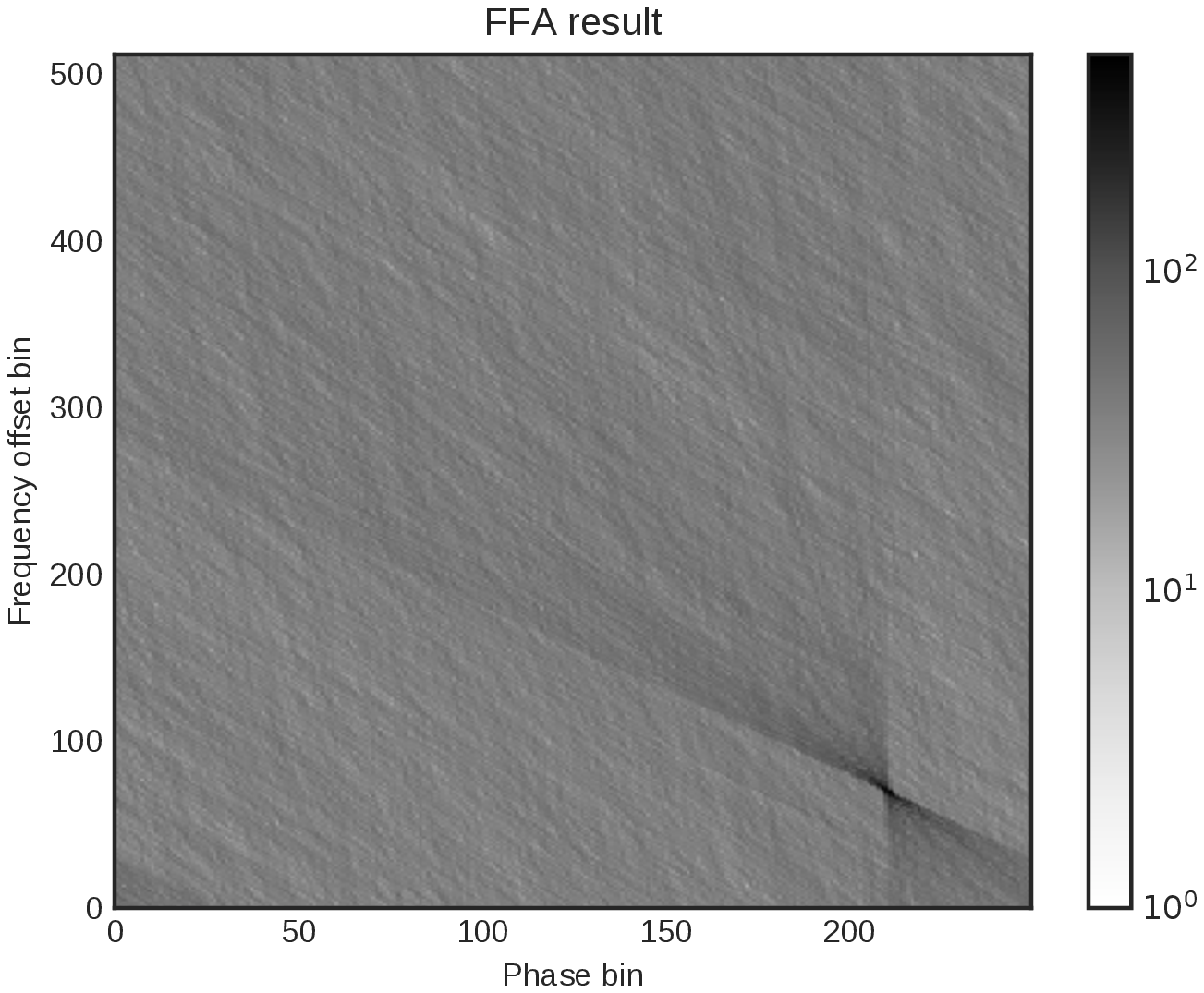}}
\caption{(Ground test data)
    a. Two-dimensional histogram input to the Fast Folding Algorithm (FFA). Photon detections are binned by time $t_i$ and fractional phase $\phi_i$ relative to the nominal period. The time bins span the full 180-second dataset, and the phase bins extend from 0 to 1. A faint line is visible, providing a preliminary indication that there is a periodic signal.
    b. Result of the FFA applied to (a). The point with the highest value corresponds to the peak frequency and phase of the input data.
    }
\label{fig:hrt-ffa}
\end{figure}

If a periodic signal is present in time series data, the fractional phase of the signal relative to the period (Equation \ref{eq:frac-phase}) has a constant value over time. If the phase is measured relative to a period which is close to but not exactly the true period, the measured phase will change linearly with time (examples are shown in Figures \ref{fig:hrt-ffa}a and \ref{fig:hrt-phase-cut}a). The FFA determines the most likely slope of this linear relationship, and therefore the true period, using a sequence of shifting and adding steps on a two-dimensional histogram of the time series data binned by time and phase.

The FFA has advantages over the more familiar Fast Fourier Transform (FFT) for the ELROI clock period recovery task. The FFA concentrates the power of the non-sinusoidal ELROI signal, while the FFT disperses power at higher harmonics into other bins. The FFA can use the known value of $T_{\textrm{nom}}$ to search for periodic signals in a specified frequency range with a specified frequency spacing, while the FFT must be calculated over all frequencies from zero to the highest frequency of interest with a fixed frequency spacing. Finally, unlike the FFT, the FFA can also be extended to measure slow changes in frequency over the duration of an observation (e.g., due to temperature drift).

A Python implementation of the FFA was used to find the true clock period in this example \cite{Petigura}. The inputs to the FFA are the photon detection times $t_i$ in bins of width $b_t$, and the photon phases $\phi_i$ relative to the nominal period $T_{\textrm{nom}}$ in bins of width $b_{\phi}$ (Figure \ref{fig:hrt-ffa}a). The optimal width of $b_{\phi}$ is the approximate width of the phase peak $\tau/T$. The width of $b_t$ is an integer number of nominal periods determined by $\Delta f$, the maximum frequency deviation to be included in the search (larger $\Delta f$ requires more timebins). Using $b_{\phi}$ = 0.004, $\Delta f$ = 0.01 Hz, and $b_t$ = 0.352 s, the FFA found a peak period of $T_{\textrm{true}} = 499.9996115$ $\mu$s in our three-minute dataset, corresponding to a peak frequency of $f_{\textrm{true}}$ = 2000.001554 Hz (Figure \ref{fig:hrt-ffa}b). The deviation from the nominal period ($T_{\textrm{nom}}$ = 500 $\mu$s) is less than 0.8 ppm, but the correction is still critical in order to apply an efficient phase cut. The FFA also determines the phase of the periodic signal in the data (relative to the true period), $\phi_{\textrm{peak}}$ = 0.16 cycles.

\subsection{Phase cut}

\begin{figure}[h]
\centering 
\subfloat[]{\includegraphics[width=0.5\textwidth]{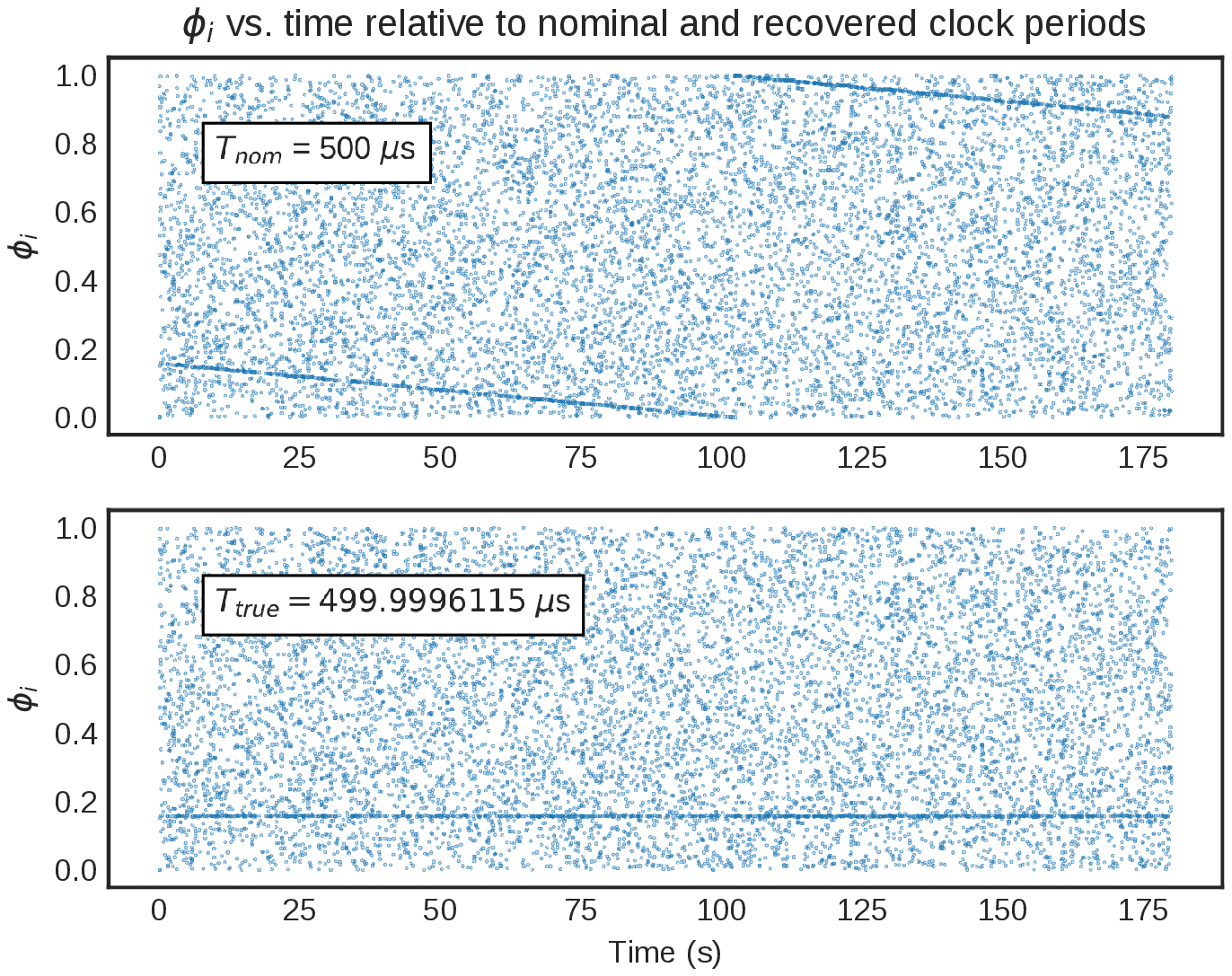}}
\subfloat[]{\includegraphics[width=0.5\textwidth]{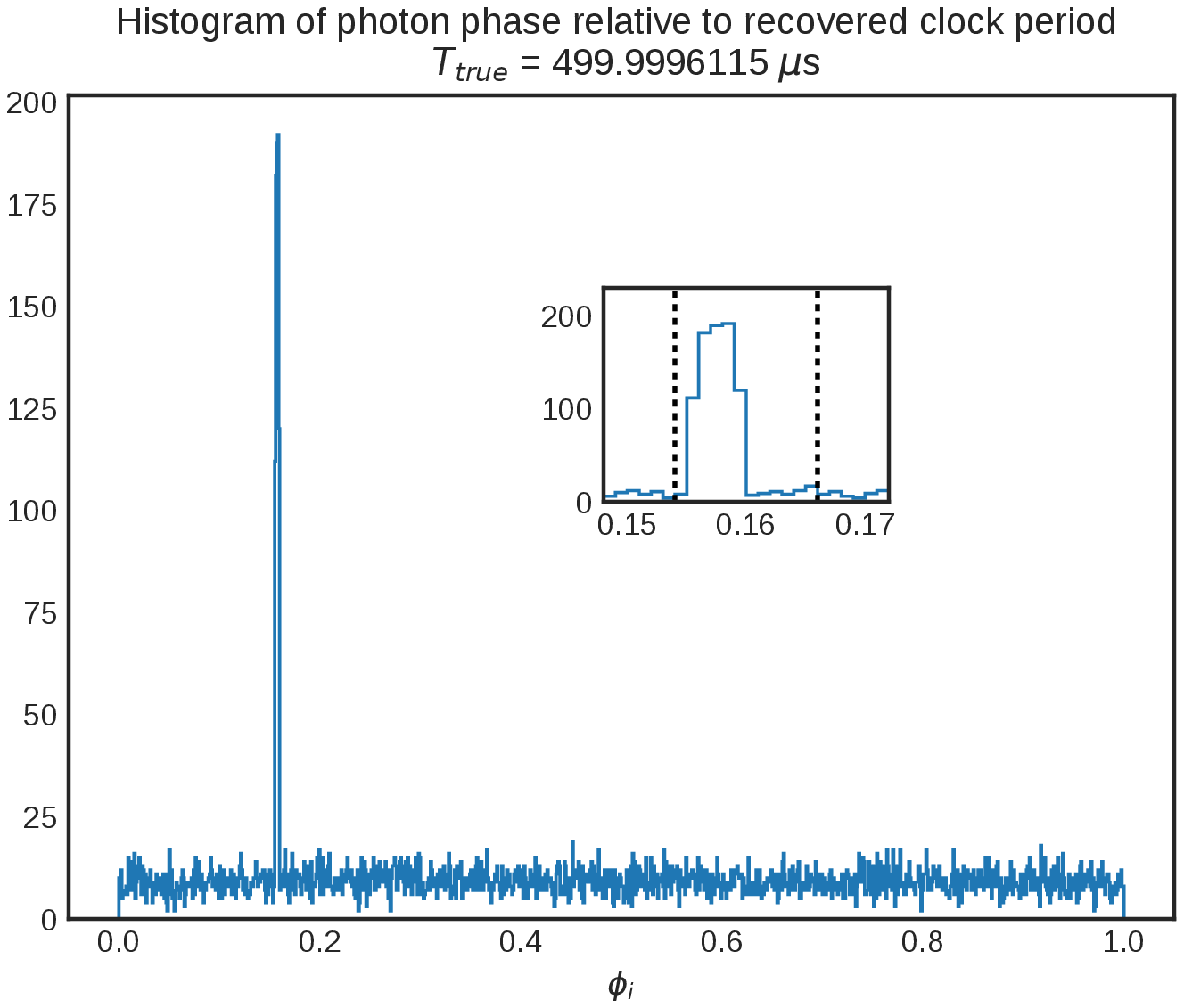}}
\caption{(Ground test data)
    a. Fractional phase $\phi_i$ of each photon detection relative to the nominal (top) and recovered (bottom) clock periods. Note that the signal phase drift over time is corrected when $\phi_i$ is measured relative to the true clock period. Photons with random phases are background detections.
    b. Histogram of $\phi_i$ relative to the true clock period $T_{\textrm{true}} = 499.9996115$ $\mu$s ($f_{\textrm{true}}$ = 2000.001554 Hz). The dashed lines in the inset show the phase cut window we will apply to this data, with a width 3 $\tau/T$. (As in Figure \ref{fig:sim-phase-cut}, each phase bin is 0.25 $\times$ $\tau/T$.)}

\label{fig:hrt-phase-cut}
\end{figure}

The true clock period, recovered using the Fast Folding Algorithm, can then be used to apply a phase cut to the photon time data. Because the FFA also determines the phase $\phi_{\textrm{peak}}$ of the clock, the phase cut can now be applied given a desired tolerance around the peak phase. We use the same tolerance as in the example of Section \ref{example-sim}, and reject all photons outside a 3 $\tau/T$ phase window (Figure \ref{fig:hrt-phase-cut}b). 

The phase cut window is not perfectly centered on the phase peak, nor is the peak as narrow and tidy as in the simulated example of Section \ref{example-sim}. The phase peak is wider than $\tau/T$, indicating that we still do not have exactly the correct clock period. The true location of the phase peak also appears slightly offset from the phase found by the FFA, which is due to the size of the phase bins $b_{\phi}$ and the uncertainty in the best period. If necessary, some additional sensitivity could be recovered by making small adjustments to $T$ and $\phi_{\textrm{peak}}$ to maximize the height of the phase peak and allow the narrowest possible phase cut. However, this step is not needed to get good results in the current example.

After the phase cut, we are left with $N_{\textrm{kept}} = 868$ photon detections, out of the original $N_{\textrm{tot}} = 9,997$ in this three-minute dataset. Therefore, we can estimate that the background photon count rate in this dataset is
\begin{equation}
R_b = \frac{N_{\textrm{tot}} - N_{\textrm{kept}}}{(1 - 3 \tau/T)t_{\textrm{obs}}} = 51.3 \textrm{ photons/s}
\end{equation}
where $3 \tau/T$ is the width of the phase cut window in this example. The estimated signal photon count rate is
\begin{equation}
R_s = \frac{N_{\textrm{kept}}}{t_{\textrm{obs}}} - (3 \tau/T)R_b = 4.2 \textrm{ photons/s}
\end{equation}
where $(3 \tau/T)R_b$ is an estimate of the rate of background photons remaining after the phase cut. We can also calculate the expected values of $N_{j'}$ for 0 bits and 1 bits after the phase cut:
\begin{equation}
\langle N_0 \rangle = \frac{(3 \tau/T)R_b \times t_{\textrm{obs}}}{m} = 0.87
\end{equation}
\begin{equation}
\langle N_1 \rangle = \langle N_0 \rangle + \frac{R_s \times t_{\textrm{obs}}}{n} = 12.7
\end{equation}

\begin{figure}[h]
\centering 
\subfloat[]{\includegraphics[width=0.5\textwidth]{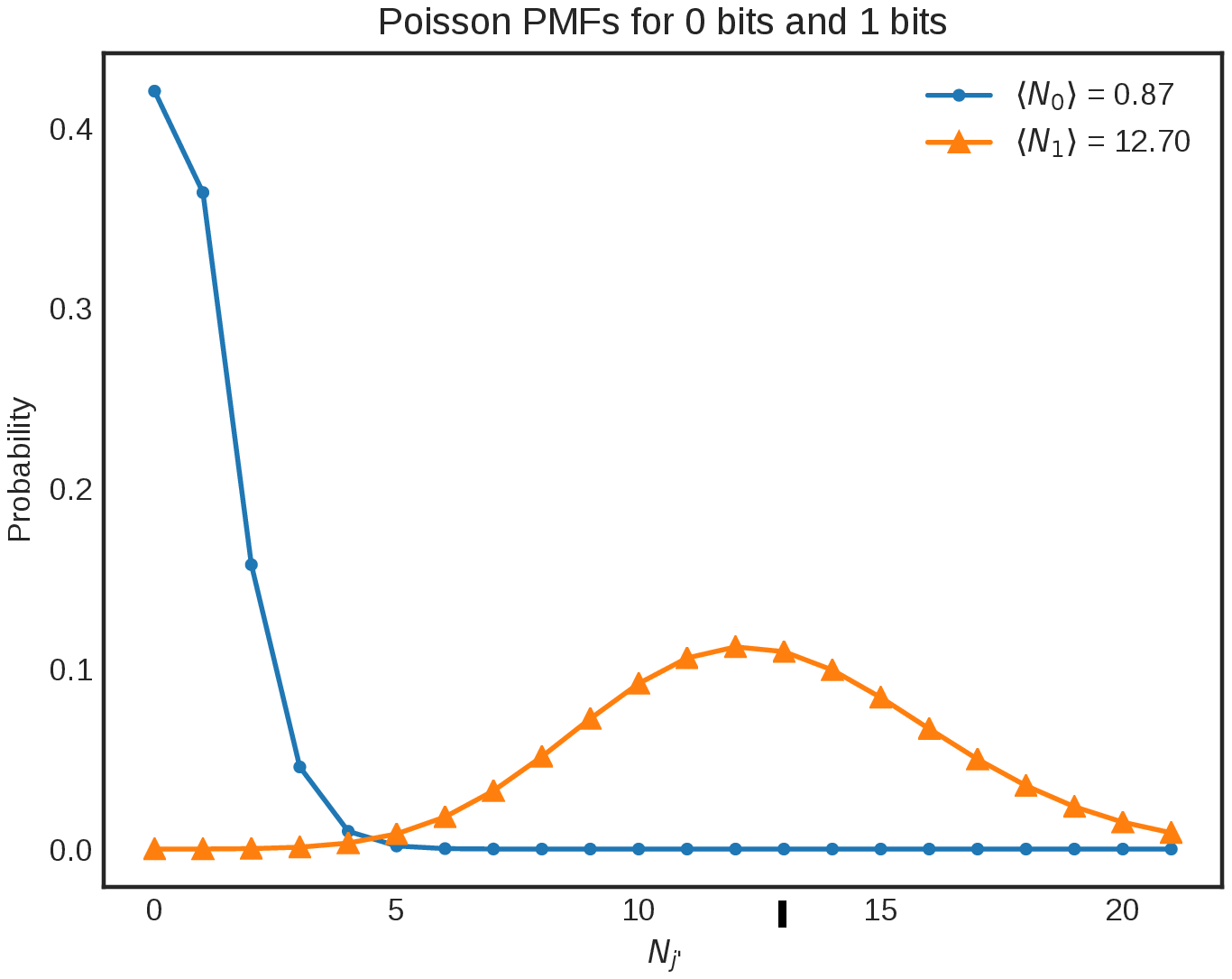}}
\subfloat[]{\includegraphics[width=0.5\textwidth]{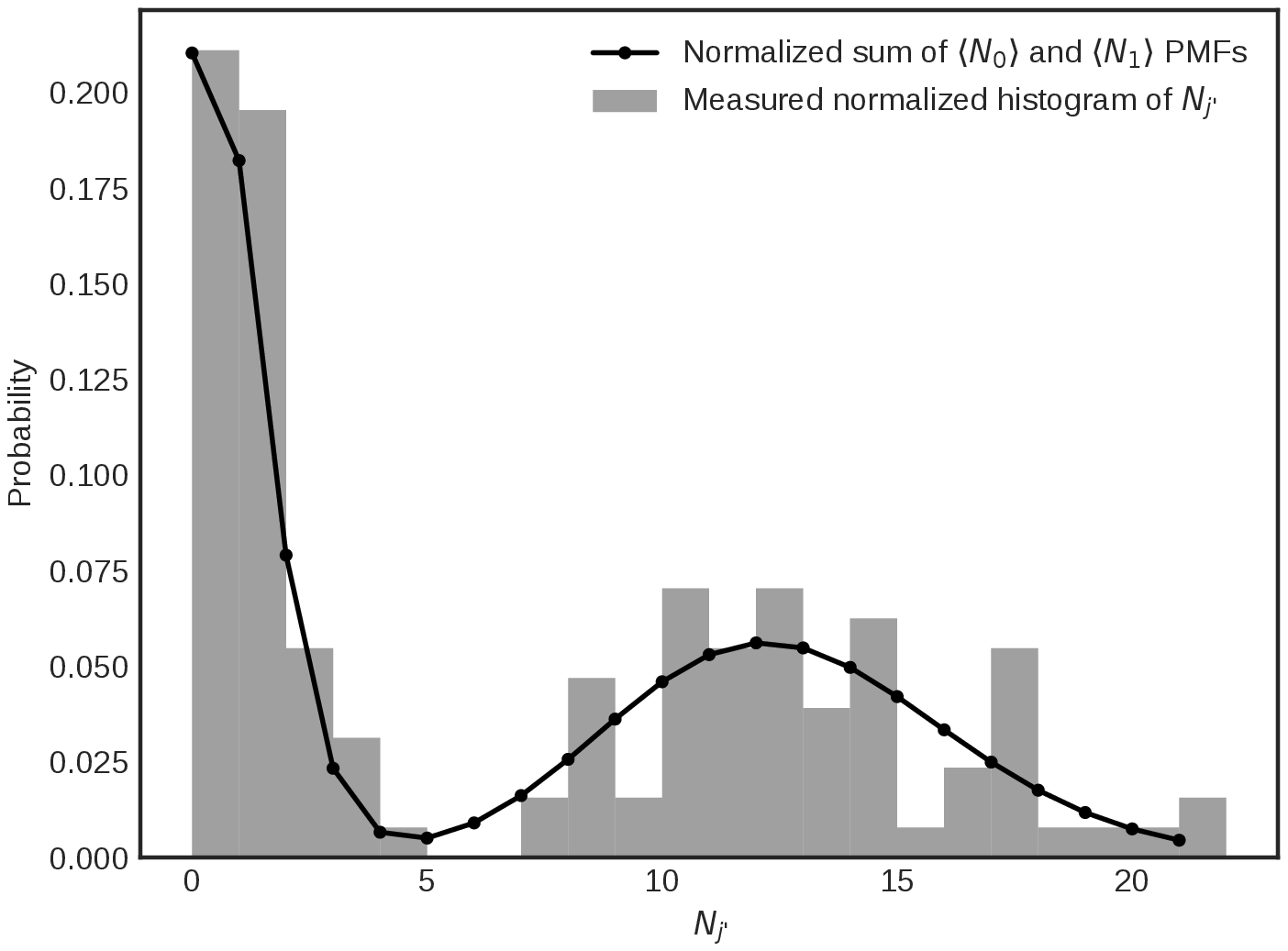}}
\caption{(Ground test data)
    a. Poisson probability mass functions (PMFs) with means equal to the measured expectation values $\langle N_0 \rangle$ and $\langle N_1 \rangle$.
    b. The measured distribution of $N_{j'}$ (for 868 detected photons in 128 bits) agrees well with the combined PMFs for 0 bits and 1 bits.}
\label{fig:hrt-pmfs}
\end{figure}

The values of $N_{j'}$ for 0 bits and 1 bits are Poisson distributed, neglecting any saturation effects in the sensor. As shown in Figure \ref{fig:hrt-pmfs}, the measured distribution of $N_{j'}$ is well described by the normalized sum of two Poisson distributions with means $\langle N_0 \rangle$ and $\langle N_1 \rangle$.

\subsection{Decoding the ID number}

\begin{figure}[h]
\centering 
\subfloat[]{\includegraphics[width=0.75\textwidth]{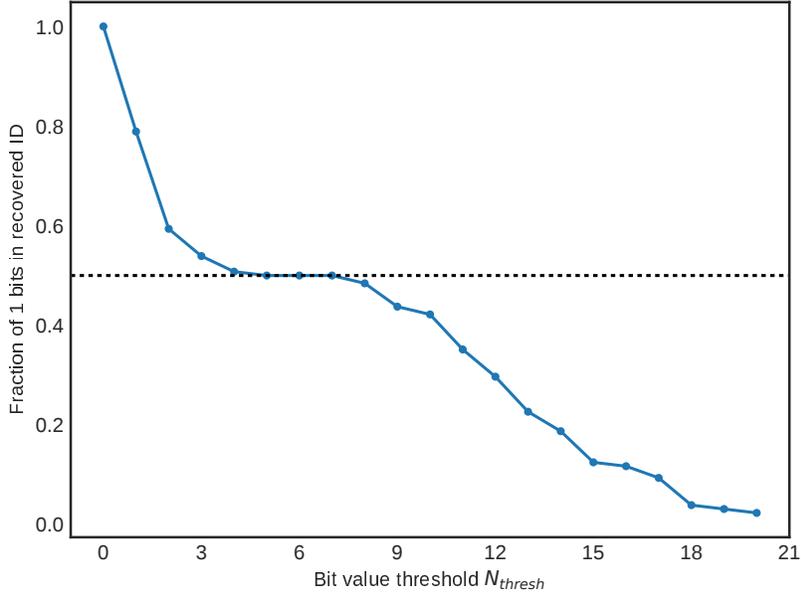}}
%\par
%\subfloat[]{\includegraphics[width=0.6\textwidth]{img/hrt-bit-errors.pdf}}
\caption{(Ground test data)
    a. The fraction of 1 bits in the recovered code as a function of the bit value threshold $N_{\textrm{thresh}}$. The known fraction of 1 bits is 0.5 for our 64-of-128-bit error-correcting code, and this information can be used to select an appropriate threshold.
%    b. Histogram of the number of bit errors in each of 20 registry IDs with each shift $0 \leq s < m$ (2560 shifted IDs total), compared to the recovered ID. The single ID with zero errors is ID number 3 in the registry (the ID number that Unit 2 was programmed to emit).
}
\label{fig:hrt-bit-errors}
\end{figure}

As in the example of Section \ref{example-sim}, the remaining photon detection times are folded by the code length $mT$ and binned by the clock period $T$ to obtain a histogram of bit photon counts (Figure \ref{fig:hrt-pmfs}b). In this example, there are again two clear bit photon count distributions corresponding to 0 bits and 1 bits. This time we will use a quantitative method to determine the best bit value threshold for hard decision decoding.

If constant-weight ID numbers are used, $N_{\textrm{thresh}}$ can be chosen to split the $b_{j^\prime}$ values into the expected number of zeros and ones. The ID number for this ground test data (and all others used in this paper) has a fraction of 1 bits $n/m$ = 0.50. As discussed further in Section \ref{low-snr}, choosing a threshold that results in $n/m$ as close as possible to the expected value minimizes the number of incorrect bits. (This method can also be extended to non-constant-weight codes if the fraction of 1 bits is restricted to some known range.) Figure \ref{fig:hrt-bit-errors}a shows the fraction of 1 bits in the recovered ID as a function of the bit value threshold. Given several thresholds that result in $n/m$ = 0.5, we arbitrarily choose the lowest one and set $N_{\textrm{thresh}} = 5$. The value of each bit is then determined using the hard decision method described in Section \ref{sim-decoding}.

Finally, we compare the recovered ID number to all the IDs in the registry (with each shift $0 \leq s < m$) to find the matching ID with the smallest Hamming distance. The recovered ID has zero errors compared to ID 3 in the registry (with a shift of 85), and indeed, ID 3 is the ID number used to program beacon Unit 2 in the ground test.

%\begin{table}
%\centering
%\caption{(Ground test data) Bit errors in the ID number recovered from a 3-minute dataset compared to 20 possible registry ID numbers generated with a 64-of-128-bit error-correcting code. The bit errors shown are for the shift $s$ that produced the minimum number of errors. ID number 3 is the ID used for beacon Unit 2 in the ELROI ground test.}
%\label{table:hrt-bit-errors}
%\begin{tabular}{l|llllllllllllllllllll}
%Registry ID & 1  & 2  & \bf{3} & 4  & 5  & 6  & 7  & 8  & 9  & 10 & 11 & 12 & 13 & 14 & 15 & 16 & 17 & 18 & 19 & 20 \\ \hline
%Bit errors  & 51 & 44 & \bf{0} & 48 & 50 & 50 & 52 & 50 & 50 & 50 & 48 & 50 & 50 & 52 & 52 & 50 & 44 & 50 & 52 & 46
%\end{tabular}
%\end{table}

\subsection{Data with reduced signal counts}
\label{low-snr}

\begin{figure}[h]
\centering 
\subfloat[]{\includegraphics[width=0.5\textwidth]{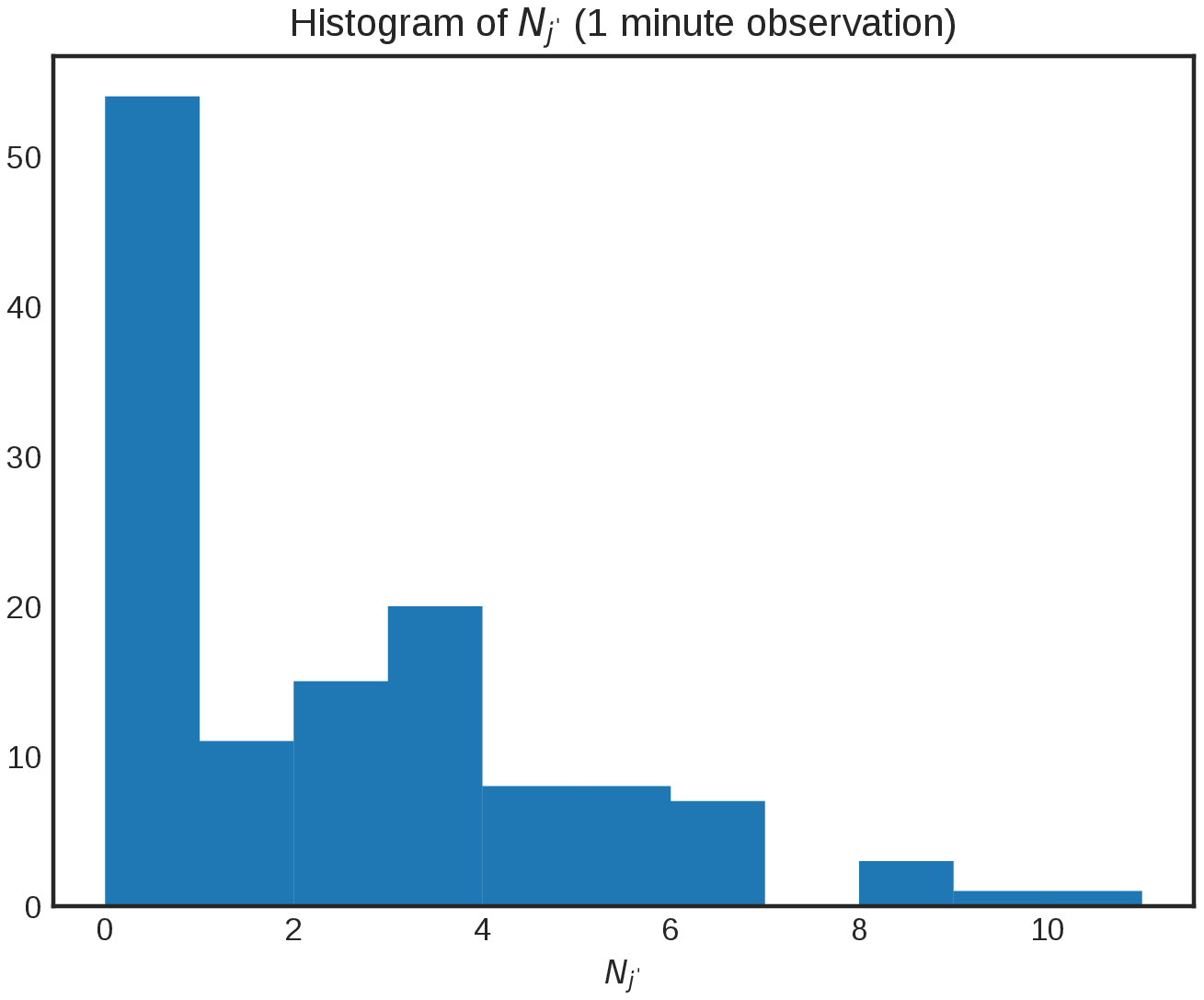}}
\subfloat[]{\includegraphics[width=0.5\textwidth]{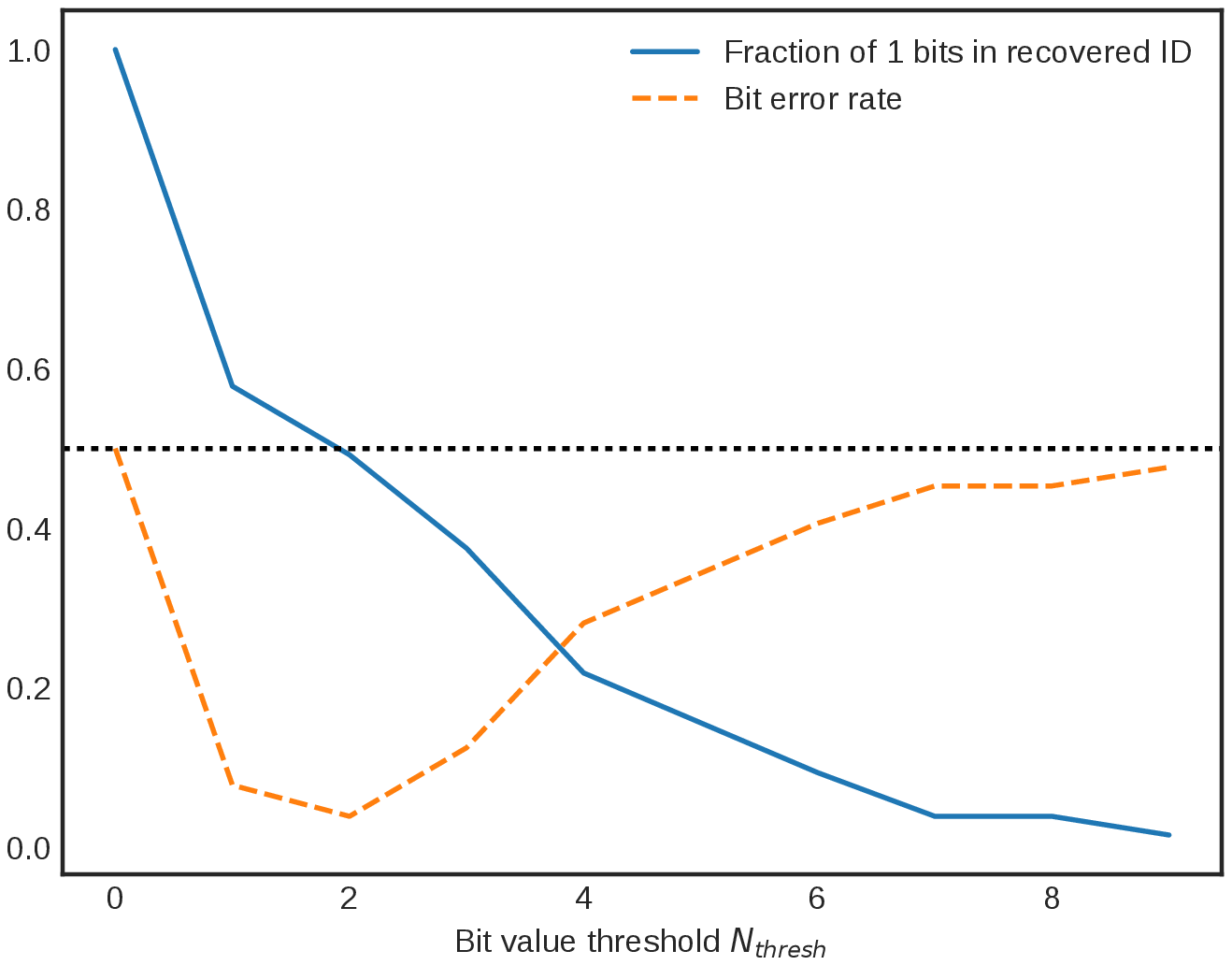}}
\caption{(Ground test data)
    a. Histogram showing the frequency of photon numbers in time bins corresponding to the bits of the code, for a 1-minute subset of the 3-minute dataset analyzed previously.
    b. In this example, there is no clear distinction between the distributions of 1 bits and 0 bits, but the known fraction of 1 bits may still be used to determine the most appropriate threshold. The bit error rate (fraction of bits that are incorrect compared to the known ID number) has a minimum at the value of $N_{\textrm{thresh}}$ that achieves the known fraction of 1 bits $n/m$ = 0.5.}
\label{fig:hrt-bit-counts-1}
\end{figure}

The previous two examples, using simulated and experimental data, both showed a clear separation between the photon number distributions of 0 bits and 1 bits, and both recovered an ID that exactly matched an ID in the registry. The signal and background rates in these two examples were both within the expected range for an ELROI beacon on a small LEO object under good observing conditions, which demonstrates that the data analysis in that case is expected to be very straightforward. But by restricting our analysis to a shorter one-minute portion of the ground test data, we can see what happens when the number of signal photons detected is so low compared to the background that the distributions of 0 bits and 1 bits significantly overlap, and bit errors become inevitable. 

The simple techniques of the previous two examples are still useful even with reduced counts. When each step in the data processing is identical to the analysis of the three-minute dataset, including recovering the clock period and phase with the FFA and applying a 3 $\tau/T$ phase cut, the resulting bit photon count histogram is shown in Figure \ref{fig:hrt-bit-counts-1}a. Although there is no longer an obvious distinction between the distributions of 0 bits and 1 bits, the fraction of 1 bits $n/m$ may still be used to choose the best value of $N_{\textrm{thresh}}$. Figure \ref{fig:hrt-bit-counts-1} shows that choosing a threshold that results in $n/m$ as close as possible to the expected value minimizes the number of incorrect bits. $N_{\textrm{thresh}} = 2$ achieves $n/m$ = 0.5 for this dataset.

The best recovered ID has five bit errors compared to the true ID. However, all other IDs have significantly more errors (the second-best match has 45 bit errors) so even using this shortened dataset, we can confidently identify the ID number of the beacon. Because of the ECC scheme used to generate the registry IDs (Section \ref{ecc-intro}), all incorrect IDs are guaranteed to be larger Hamming distance from the recovered ID than the true ID, up to 12 bit errors in the recovered ID. As discussed in Sections \ref{error-rates} and \ref{conclusion}, more sophisticated analysis techniques, improved ECC schemes, and operational context may be able to further improve our confidence that the recovered ID number is correct even for $>12$ bit errors.

\section{Error rates}
\label{error-rates}

\begin{figure}[h]
\centering 
\subfloat[]{\includegraphics[width=0.5\textwidth]{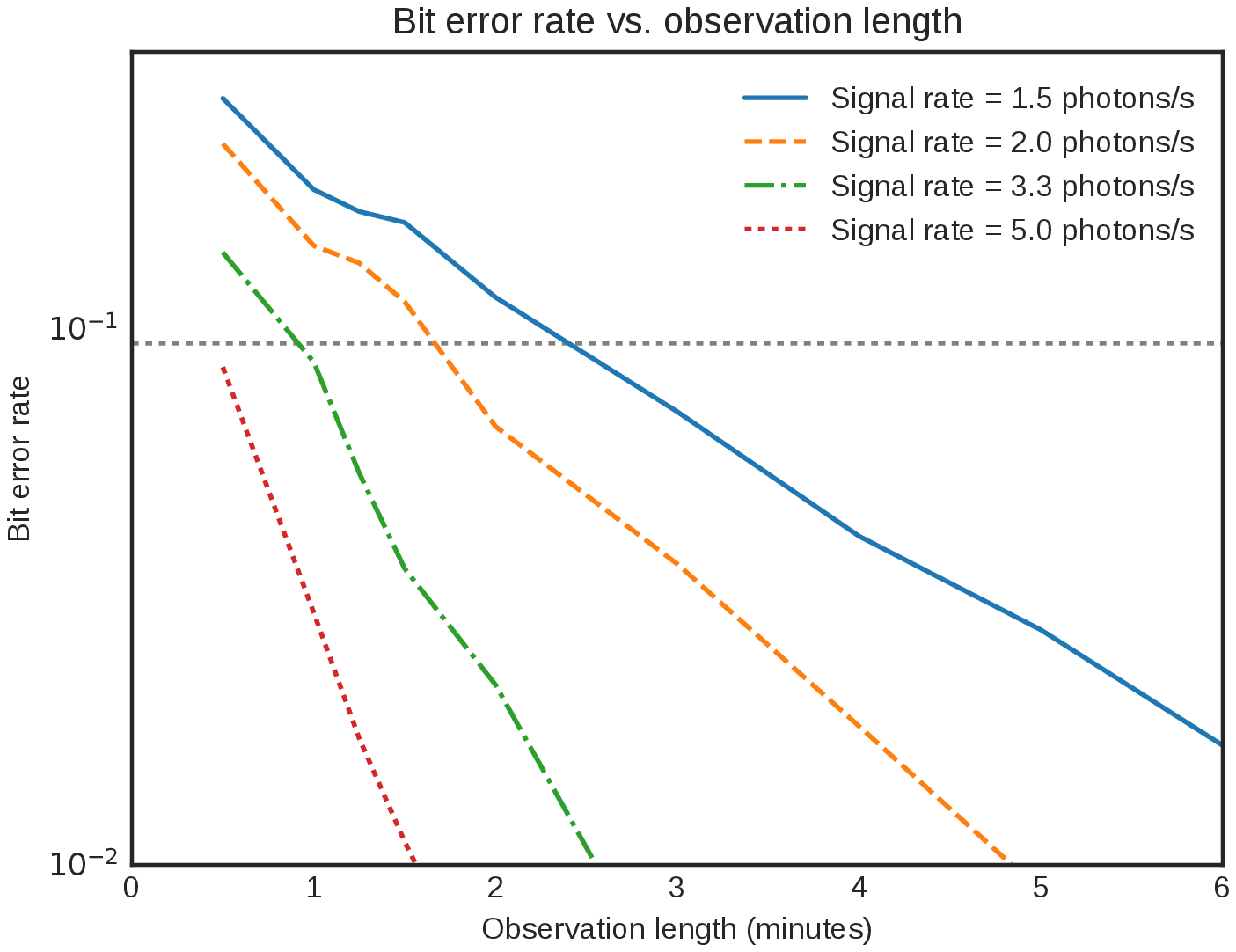}}
\subfloat[]{\includegraphics[width=0.5\textwidth]{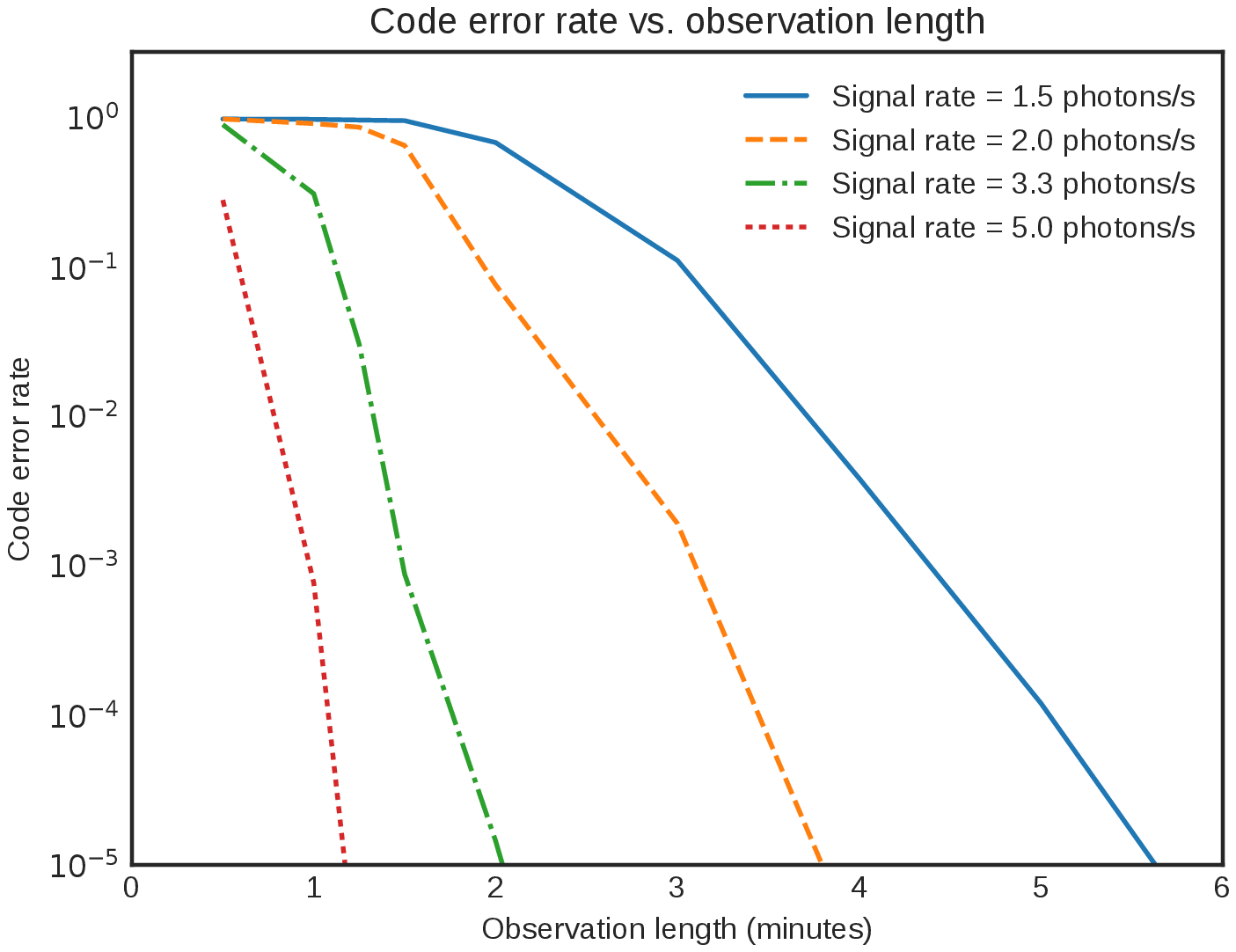}}
\caption{Numerical simulation of bit error rates and codeword error rates in recovered ELROI IDs, for different signal rates (in photons/s).  The background rate for all observations (after spectral filtering and phase cut) is assumed to be 0.36 photons/s, an estimated value for a small sunlit LEO satellite. The signal rate of 3.3 photons/s is the value estimated for a typical beacon at 1000 km range in \cite{Palmer2018}. Each data point represents $10^7$ simulated observations.
    a. Simulated bit error rates. The dashed horizontal line indicates a bit error rate of 12/128.
    b. Simulated codeword error rates.}
\label{fig:cer}
\end{figure}

As discussed in Section \ref{ecc-intro}, all the example ELROI IDs used in this paper are 128-bit binary sequences with a minimum 24-bit distance between a given ID and all other sequences (allowing cyclic permutation). This guarantees that for up to 12 incorrect bits, the true ID will be closer to the recovered ID than any other ID in the ELROI registry. Thus, if the bit error rate (frequency of individual bit errors) in a given recovered ID is 12/128 or lower, the codeword error rate (probability of incorrect beacon identifications) for that ID is zero. For more than 12 bit errors, the codeword error rate increases with the number of bit errors.

Figure \ref{fig:cer} shows the results of a numerical simulation of bit error rates and codeword error rates in recovered IDs. Bit photon counts $N_{j}$ were randomly generated for 0 bits and 1 bits, and the hard decision decoding method described in Section \ref{low-snr} was used to determine bit values, with the threshold chosen to achieve the fraction of 1 bits closest to 0.50 for each recovered ID. If the number of bit errors in an ID was 13 or more, it was counted as a codeword error. This is a conservative approach, as a recovered ID with more than 12 bit errors may still be closer to the true ID than any other ID in the registry.

The error rates in Figure \ref{fig:cer} may appear high compared to typical optical communications applications, but they are acceptable for the ELROI application. Note that for a signal rate of 3.3 photons/s (comparable to our current LEO prototypes) the code error rate is 1 in 100,000 after only two minutes of observation. This means that the chance of mis-identification is 1 in 100,000 after two minutes if all ID numbers in the ECC scheme are in use. In practice, context will also give additional information, and not all possible IDs in the registry will be equally likely---for example, if 100 CubeSats are launched together into a known orbit, all carrying ELROI beacons, the codeword error rate for identifying one of these CubeSats will be greatly reduced compared to a complete registry of thousands or millions of IDs. Additional observation time also reduces the chance of mis-identification.

\begin{figure}[h]
\centering
\includegraphics[width=0.75\textwidth]{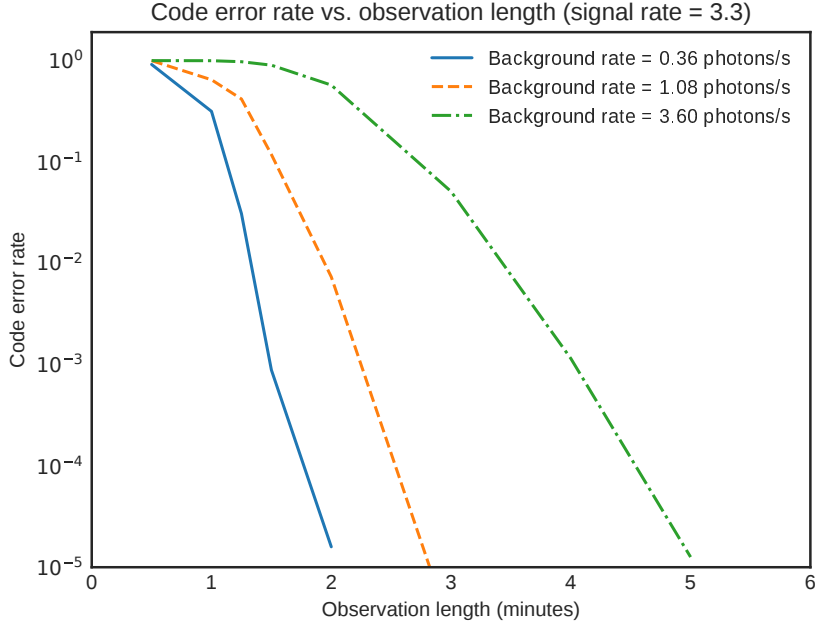}
\caption{Numerical simulation of codeword error rates with a fixed signal rate of 3.3 photons/s and three different background rates (after spectral filtering and phase cut). 0.36 background photons/s is a reasonable estimate for a LEO CubeSat.}
\label{fig:cer-bg}
\end{figure}

According to our link budget calculations, signal rates of 3 to 20 photons/s will be typical at our ground station for prototype ELROI beacons on LEO satellites, depending on the beacon power and range\cite{Palmer2018,Holmes2018}. We expect to be able to identify our current flight prototypes with codeword error rates of $10^{-5}$ or lower, and under ideal conditions codeword error rates may reach $10^{-9}$.

Figure \ref{fig:cer-bg} shows how codeword error rates vary with the background rate for a fixed signal rate of 3.3 photons/s. If the actual observed background rate is three times higher than expected, less than a minute of additional observation is needed to reach a codeword error rate of $10^{-5}$. If the observed background rate is ten times higher than expected, approximately three minutes of additional observation are needed to reach the same codeword error rate. In this case, the ID may not be able to be identified with the same level of confidence in a single pass over a ground station, but can still be identified with lower confidence. Higher-power beacons may be used to improve performance for large satellites with high background rates from reflected sunlight (further discussion of these design trade-offs may be found in Reference~\citenum{Palmer2018}).

\section{Conclusions}
\label{conclusion}

We have described the basic data analysis techniques needed to recover the unique ELROI ID number for a space object, and we invite readers to consider observing our upcoming test flights.
% We invite readers to experiment with these techniques themselves using the code and sample data available at \url{http://github.com/XX}, and by making their own observations of upcoming ELROI test flights.
As we have shown with both simulated and ground test data, even the simple methods described here can easily recover the beacon ID for a small LEO satellite.

However, further optimization is possible and should be explored. Improvements to both the encoding and decoding schemes may be considered. The examples shown here have used a hard decision decoding method. Although Figure \ref{fig:hrt-bit-counts-1} suggests this method may be close to optimal, a registry search for the best ID after hard decision decoding weights all bit errors between the recovered ID and a registry ID equally, without considering the actual number of photons detected for each bit. A soft decision decoding approach might improve on this by weighting bit errors by the number of photons detected.

Although the simple ECC scheme presented here is sufficient for many practical use cases, there are codes that achieve substantially better Hamming distance (and therefore improved error tolerance) for the same code length and number of codes; for example, BCH codes \cite{Hocquenghem1959,Bose1960}. There may also be an advantage to using forward error correction and Bayesan decoding with prior information, as implemented in (for example) turbo codes, LDPC codes, and, recently, polar codes \cite{Arikan2009,Gallager1962,Berrou1993}. These types of ECC codes are frequently used in modern digital communications, including space communications, and have known optimal decoding techniques.

\section*{Funding}

US Department of Energy through the Los Alamos National Laboratory (LANL) Laboratory Directed Research and Development (LDRD) program\\ \\
Richard P. Feynman Center for Innovation at Los Alamos National Laboratory\\ \\
Center for Space and Earth Sciences at Los Alamos National Laboratory

\section*{Acknowledgments}
ELROI hardware and software was developed and tested at LANL by Louis Borges, Richard Dutch, Darren Harvey, Ryan Hemphill, David Hemsing, Alexandra Hickey, Lee Holguin, Zachary Kennison, Jim Lake, Joellen Lansford, and Charles Weaver. The ELROI long-range ground test was assisted by Amanda Graff, David Graff, Michael Rabin, and David Thompson.

%%%%%%%%%%%%%%%%%%%%%%% References %%%%%%%%%%%%%%%%%%%%%%%%%

%%%%%%%%%% If using BibTeX:
%\bibliographystyle{ieeetr}
\bibliography{ref.bib}

\end{document}